# THE GREAT JANUARY COMET OF 1910 (C/1910 A1): A KEY OPPORTUNITY MISSED BY NEW ZEALAND ASTRONOMERS

**John Drummond**
*University of Southern Queensland, West Street, Toowoomba, QLD 4350, Australia.*
E-mail: kiwiastronomer@gmail.com

**Wayne Orchiston**
*University of Science and Technology of China, Hefei, Anhui, China; and University of Southern Queensland, West Street, Toowoomba, QLD 4350, Australia.*
E-mail: wayne.orchiston@gmail.com

**and**

**Carolyn Brown and Jonathan Horner**
*University of Southern Queensland, West Street, Toowoomba, QLD 4350, Australia.*
E-mails: Carolyn.Brown@unisq.edu.au
jonti.horner@unisq.edu.au

**Abstract:** C/1910 A1 was one of the Great Comets of the twentieth century. Although it was widely observed from the Northern Hemisphere, it was first discovered by observers south of the Equator. The comet arrived just months before the widely anticipated apparition of Comet 1P/Halley and was significantly more spectacular. As a result, the two comets were confused, and many who, in later years, talked about how prominent Comet 1P/Halley was in 1910 were often remembering C/1910 A1. In this paper, we present the results of a detailed search through historical records and media publications in Aotearoa / New Zealand, to investigate how extensively C/1910 A1 was observed from New Zealand. We compare our results with observations reported for Comet 1P/Halley later in 1910, finding that surprisingly few observations of C/1910 A1 were made by New Zealand observers. We discuss cases where the comet was misidentified as being an early sighting of 1P/Halley and compare the observations made in New Zealand with international observations/records/accounts. We find that, although the Great January Comet of 1910 was observed from New Zealand, it was witnessed by few compared to other parts of the world, meaning that the apparition of C/1910 A1 was something of a missed opportunity for New Zealand astronomers.

**Keywords:** comets; New Zealand; C/1910 A1 (the Great January Comet); 1P/Halley; newspaper reports.

## 1 INTRODUCTION

Two bright naked eye comets appeared in the sky in 1910 (CE). One was expected, the other was not. Anticipation of the apparition of Halley's Comet (henceforth Comet 1P/Halley—unless in a quote) was building around the world through books (e.g. Chambers, 1909: 124–125; Elson, 1910: 63–75; Guillemin, 1877: 107), newspaper articles (e.g. *Auckland Star*, 1910; Amusements, 1910; Halley's Comet Approaching, 1908), and public lectures (e.g. *New Zealand Herald*, 1909: 5; *Northern Advocate*, 1910: 5; *Town and Country*, 1910: 6). Some recalled the 1835 apparition of 1P/Halley with glowing accounts, stating it was a "… magnificent object …" (Smith, 1986: 2) with a 20° tail (A Famous Comet, 1894). Many wanted to view the most famous of all comets for themselves. However, the appearance of a bright comet in January 1910 (Rolston, 1910: 373–1374) caused considerable confusion for the public, and many misidentified it as 1P/Halley, which would reach peak brightness some months later. With the general public eagerly awaiting the appearance of Comet 1P/Halley in 1910, it was only natural that many would believe that the bright comet they saw earlier that year was 1P/Halley—particularly given that it was almost a decade since the last spectacular naked eye comet had been visible (C/1901 G1, Viscara) (Comet a 1901: 143; Kronk, 2007: 10–114). This was particularly true given that astronomers had been spoiled during the previous century with the appearance of an unusually large number of spectacular bright comets—with four major naked eye comets being visible in just the 1880s alone (see e.g. Drummond, 2023; Kronk, 2003; Orchiston et al, 2020a).

This paper focuses on the 'Great January Comet' (Rolston, 1910: 372), referred to here as C/1910 A1. We compare observations made of this comet from Aotearoa / New Zealand[1] with international observations and note the similarities and differences. The misidentification of Comet C/1910 A1 with Comet 1P/Halley will also be addressed, and we compare the unique astrophysical traits of these two comets.

To find observations of Comet C/1910 A1 from New Zealand we used Papers Past, which we discuss below, to search New Zealand newspaper articles using the keyword 'Comet'. The search dates were from 12 January to 1 March 1910. In all, 871 articles were found, although some of these were duplicate reports, since there were a number of cases where one newspaper repeated another newspaper's content (or their common source). Although numerous articles related to Comet 1P/Halley, some did contain observations of Comet C/1910 A1 from New Zealand, including dates, locations, and observational notes.

In Section 2, we discuss the context of the apparition of Comet C/1910 A1, when anticiption of 1P/Halley had been building in the years before it became a naked eye spectacle in 1910. In Section 3, we present an overview of the physical and orbital properties of the two comets at their 1910 apparitions. Section 4 reviews the discovery and early development of Comet C/1910 A1, before we compare observations made of the comet in New Zealand with those made elsewhere in Section 5. Finally, we present our concluding thoughts in Section 6.

## 2 THE ANTICIPATION BUILDS

The two most comprehensive books with historical international comet observations are S.K. Vsekhsvyatskii's *Physical Characteristics of Comets* (Vsekhsvyatskii, 1964) and Gary Kronk's six volume *Cometography* (Kronk, 2000; 2003; 2007; 2009; Kronk and Meyer, 2010; Kronk et al., 2017). In regards to 1910 observations of Comets C/1910 A1 and 1P/Halley from New Zealand, none are found in these books. Unfortunately, due to the isolation of New Zealand in the South Pacific and time delays of ship-born mail, comet observations, drawings and photographs were infrequently submitted to overseas observatories or journals (Orchiston, 1999: 212–221). To send a telegram to England cost the equivalent of ~NZ$120 in today's currency (for New Zealand's isolation and problems with submitting observations overseas see Orchiston et al., 2020b: 659–674). For these reasons, comet observations from New Zealand in international publications are rare, and no observations of either C/1910 A1 or the 1910 apparition of 1P/Halley from New Zealand can be found in the works of Kronk or Vsekhsvyatskii. Instead, many observers simply reported their observations in local or national newspapers as a method of disseminating information about a comet and the observations they made.

In order to study reports of Comets C/1910 A1 and 1P/Halley in the New Zealand press of the time, we make use of the public archive of that data, made available through Papers Past, an online digital archive and research tool from the National Library of New Zealand. It allows users to search through a database of archived articles from the press in New Zealand, with a large catalogue of digitised historical resources available to users.[2]

All of the newspaper reports in Papers Past (https://paperspast.natlib.govt.nz/) use New Zealand Standard Time (NZST). New Zealand lies between 167° and 178° east of the Greenwich meridian; therefore it is 12 hours ahead of Universal Time (UT). Kronk (2007) and Vsekhsvyatskii (1964) use Universal Time in their books. In order to reduce possible confusion with dates, this paper, where possible, states the time in NZST first with UT placed in parentheses.

To illustrate the growing public anticipation of Comet 1P/Halley, Figure 1 shows the number of newspaper articles published about its apparition in New Zealand between January 1908 and December 1912 (based on data available in Papers Past). The graph shows that when Maximilian Wolf (1863–1932; Knill, 2014) detected 1P/Halley in September 1909 (discussed below), media interest rose dramatically. Then, when the comet reached perihelion and was at its brightest, the number of newspaper articles about 1P/Halley peaked.

As Figure 1 illustrates, public anticipation of Comet 1P/Halley increased from September 1909. Many of those who saw Comet C/1910 A1 in January 1910 assumed that it was Comet 1P/Halley. What of these two comets? Were their displays and characteristics similar? Section 3 compares and contrasts the two.

## 3 OVERVIEWS OF COMETS C/1910 A1 AND 1P/HALLEY

Before examining the reported sightings of Comet C/1910 A1 from New Zealand and comparing them with international observations during the same epoch, we compare and contrast the coinciding comets, C/1910 A1 and 1P/Halley, by looking at their orbital elements (Table 1) during their 1910 apparitions, then briefly summarising the appearance of each as recorded by observers around the world.

### 3.1 Comet C/1910 A1

C/1910 A1 was the first comet to be seen without optical aid in 1910 (Boss, 1910: 70; Kronk, 2007: 170). Astronomers described it as being "... among the great comets of history …" (The New Comet (1910a), 1910: 440) and "… weirdly magnificent." (Rolston, 1910: 372). Six diamond

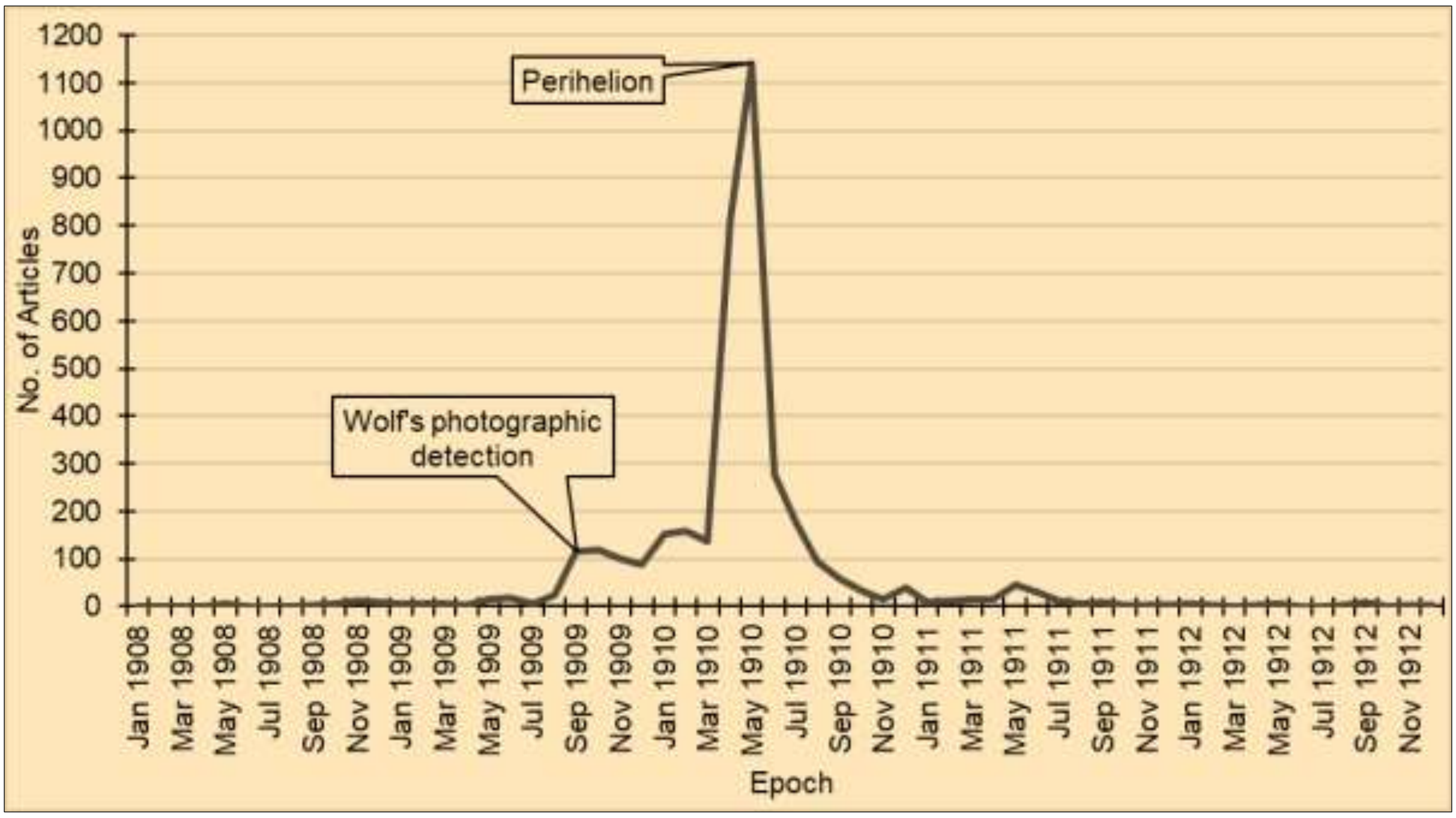

Figure 1: Newspaper articles relating to Comet 1P/Halley published in New Zealand between January 1908 and December 1912 retrieved from the Papers Past search engine. Note the rise in newspaper articles after Wolf recovered 1P/Halley and how these publications peaked at 1P/Halley's perihelion. Comet C/1910 A1 was visible to the naked eye from New Zealand between mid-January and late-January 1910 (plot: the authors).

Table 1: A comparison of the orbital elements of Comets C/1910 A1 and 1P/Halley (1910 apparition). The information provided is based on the NASA–JPL Horizons data system and Kronk (2007).

| Orbital Elements (base on NASA–JPL) | C/1910 A1 | 1P/Halley |
|---|---|---|
| Date of Perihelion | 17 January 1910 UT JD 2418705.5 | 20 April 1910 UT JD 2418800.5 |
| Period of orbit | ~14,000 years | 76.105 years |
| Perihelion distance | 0.12896 au | 0.58721 au |
| Aphelion distance | ~1200 au | 35.331 au |
| Date of Perigee | 18 January 1910 UT | 20 May 1910 UT |
| Perigee distance | 0.8581 au | 0.1514 au |
| Semimajor axis | ~590 au | 17.959 au |
| Eccentricity (*e*) | 0.99978 | 0.96730 |
| Inclination of the orbit (*i*) | 138.78° | 162.22° |
| Argument of perihelion | 320.89° | 111.74° |
| Longitude of the ascending node | 90.03° | 58.56° |

Note: Astronomers measure distances in the Solar System in 'au'. One au is one astronomical unit, which is defined as the average distance between the Earth and the Sun, namely 149,597,870 km. The elements in the table above were obtained using the NASA Horizons data system (https://ssd.jpl.nasa.gov/horizons/; accessed on 1 September 2025) and Kronk (2007: 170, 178), and are the instantaneous osculating elements at the date of perihelion for each comet. Whilst the elements from the database are given to more significant figures than presented here, in the table above we have chosen to round all values to a meaningful number of significant figures for internal consistency, with the orbital period, semi-major axis, and aphelion distance given to just two significant figures for C/1910 A1—those values are much less certain, as a very small change in the observed orbital eccentricity leads to a very large change in the calculated values of these variables.

miners at the Premier Diamond Mine in the Transvaal, South Africa, are credited with first observing Comet C/1910 A1 (Innes, 1910: 11–12; Kronk, 2007: 170; Mobberley, 2011: 53). The date of that observation is reported as either 12 or 13 January 1910 (UT) (for an example of 12 January see Vsekhsvyatskii (1964: 380) and for 13 January see Innes (1910: 11–12) and Kronk (2007: 170)). In this work, we follow Kronk (2007: 170) and consider that those discovery observations were made before dawn on 13 January NZST (13 January UT). At that time, the comet was already visible to the naked eye (Boss, 1910: 70; Pickering, 1910: 1), with an apparent magnitude of –1 (Bortle, 1998), comparable to the two brightest stars in the night sky, Sirius (magnitude –1.46) and Canopus (magnitude –0.74). Innes (1910:

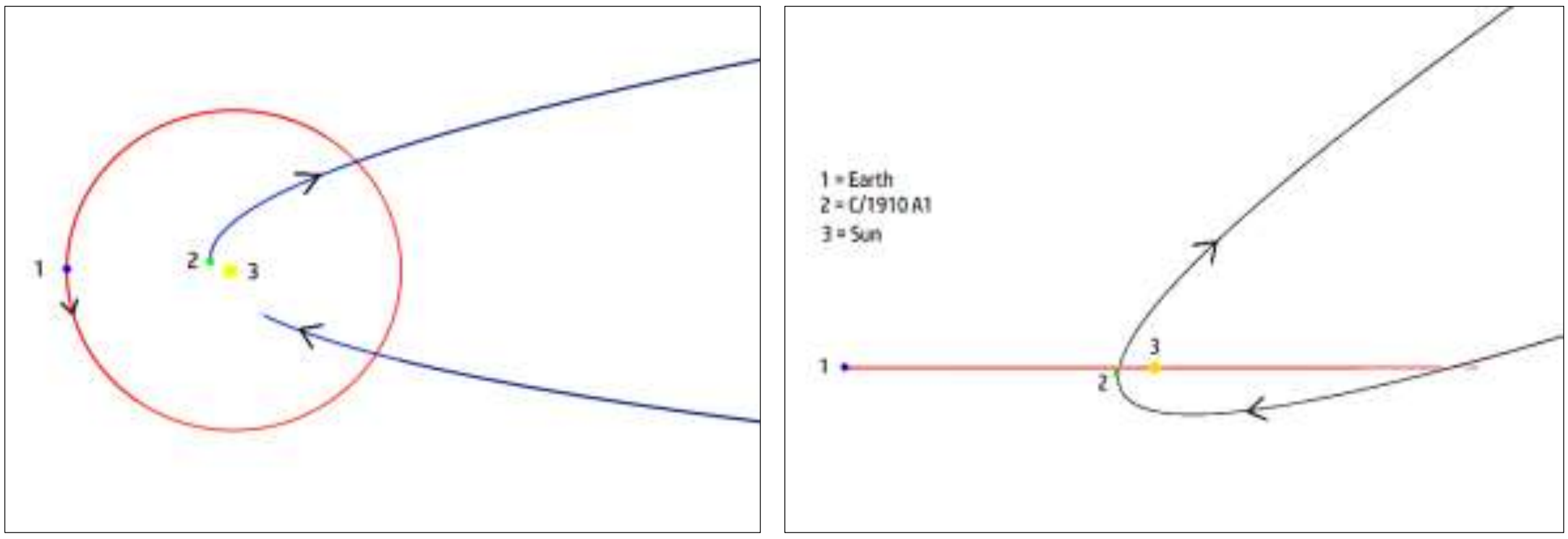

Figures 2a (left) and 2b (right): The orbit of Comet C/1910 A1 from above Earth's North Pole (2a) and side-on to Earth's orbit (2b). The locations of the Earth and the comet are set for the comet's perigee date of 18 January 1910 UT. The Earth's (1) orbit is red and C/1910 A1's (2) orbit is blue. The Sun (3) is at the centre. Note how C/1910 A1 approaches from the north, travelled under the Earth's orbital plane on the other side of the Sun and then became visible in Earth's southern hemisphere as it tracked away from the Sun and increased in elongation. The arrows show the trajectory of C/1910 A1 and Earth (courtesy: NASA/JPL–Caltech). Please note that the software removed the part of the orbit when it is not viewed through the ecliptic from the above perspective.

11) wrote that the diamond mine workers "… mistook the new comet for Halley's." Note the immediate misidentification between Comets C/1910 A1 and 1P/ Halley. That initial misunderstanding would go on to be widely repeated all over the world.

Comet C/1910 A1 approached the inner Solar System from Earth's north, passed through the plane of the ecliptic on ~9 December 1909 (UT), and was in the Earth's southern sky at its brightest (at perihelion and perigee) on 17 and 18 January 1910 respectively (NZST/UT), at a distance of about 0.8581 au (Kronk, 2007: 170). Levy (1994: 246) states that the comet was "... under wraps …" and appeared suddenly due to it "... approaching the sun from the opposite direction from us." It once again crossed the ecliptic plane heading north on ~19 January 1910 (UT). Refer to Figures 2a and 2b which display the comet's path in blue in relation to the Earth's orbit in red.

At the time of its discovery, Comet C/1910 A1 was a bright naked eye comet rising about an hour before the Sun (Innes, 1910: 11). The comet was sufficiently bright that, on the mornings immediately following its discovery, it was widely observed internationally by "... crowds …" (Rolston, 1910: 372). As it moved around the Sun, the comet shifted from the morning to the evening sky, and on 17 January (UT) reached perihelion (Curtiss, 1910a: 94; 1910b: 102; Hussey, 1910: 78). On that day it was visible in the daytime sky without optical aid (Pickering, 1910: 1; Rolston, 1910: 372), and it remained a daytime comet for an additional three days (Kronk, 2007: 171; Seargent, 2009: 144; Vsekhsvyatskii, 1964: 380) due to the forward scattering of light from the significant amounts of dust being ejected by the comet—the Sun illuminating the cometary dust from behind as seen from Earth. It was at least magnitude –4 or –5 (Bortle, 1998). Immediately after its daytime apparition during 17–20 January (UT) (Kronk, 2007: 171–172), the comet dimmed due to its increasing distance from the Earth and Sun, and as a result of the reduced effect of the forward scattering of light (which is only efficient for comets that are relatively close to the Sun in the sky). By mid-February 1910 observation of Comet C/1910 A1 required optical aid (Bortle, 1998; Kronk, 2007: 175). As it increased in elongation at night the comet's tail was observed to extend for 50° (*ibid.*).

## 3.2 Comet 1P/Halley

1P/Halley's perihelion date lagged behind that of Comet C/1910 A1 by three months; 20 April 1910 UT (Kronk, 2007: 149; Yeomans, 2007) compared to 17 January 1910 UT (Kronk, 2007: 171), respectively. When C/1910 A1 was exceptionally bright, 1P/Halley was still invisible to the naked eye (in late January 1910 it was still between magnitudes 9 and 10 (Kronk, 2007: 146; Seargent, 2009: 55), placing it between 10 and 25 times too faint to see without optical aid). Indeed, Comet 1P/Halley remained too dim for naked eye visibility until about three weeks after Comet C/1910 A1 was most visible. Comet 1P/Halley was first identified on a photograph taken by Maximilian Wolf on 12 September 1909 UT (Kronk, 2007: 141; Wolf, 1909: 179)—an observation that verified the predictions made in late 1908 by the noted British amateur astronomer William F. Denning (1848–1931; Beech, 2023), who correctly predicted that Comet 1P/Halley would be first detected at

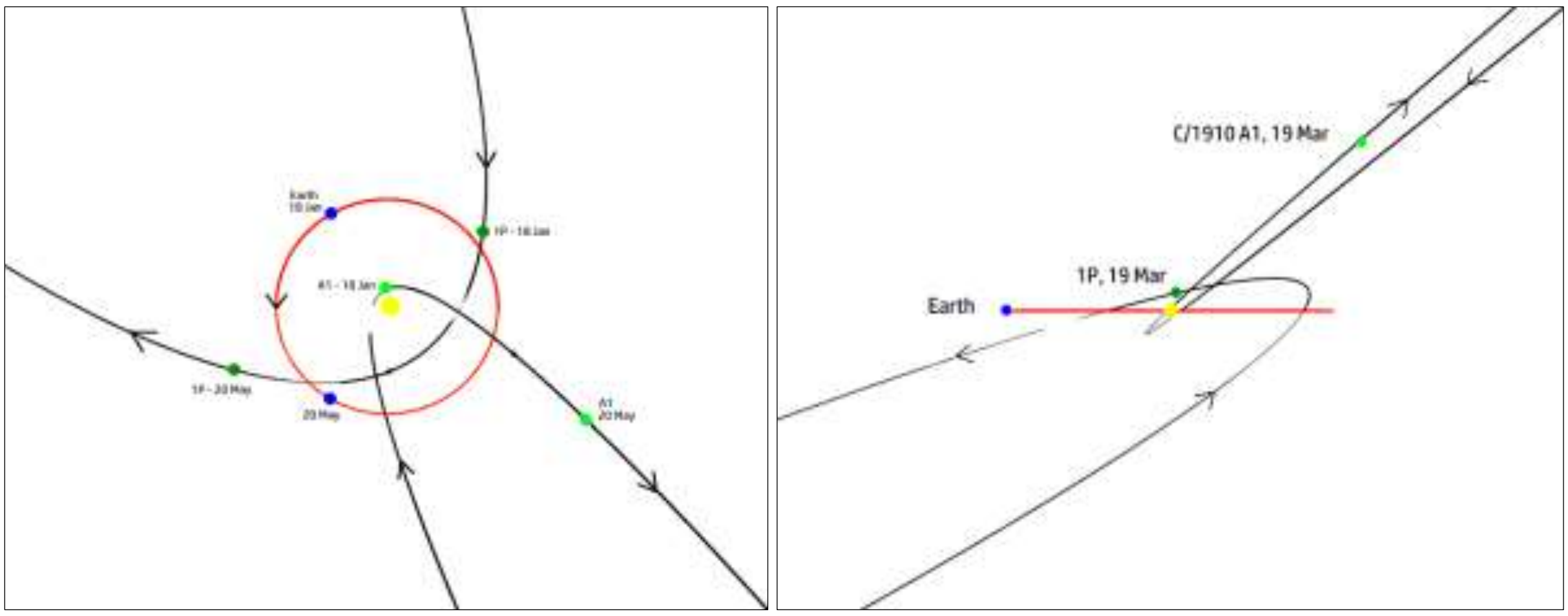

Figures 3a (left) and 3b (right): The orbits of Comets C/1910 A1 and 1P/Halley as viewed from above Earth's North Pole (3a) and side-on to Earth's orbit (3b). In 3b, Earth's (blue dot) location is set for the date between the two comets' perihelia, namely 19 March 1910. The Earth's orbit is red, and the comets on 19 May (UT) are the light green dot for C/1910 A1 and dark green dot for 1P/Halley. The orbital paths of the two comets are black. The Sun is at the centre. Note the different approach directions of the two comets (courtesy: NASA/JPL–Caltech).

its forthcoming apparition by photography, in either September or October 1909 (Denning, 1909: 62). The first international naked eye observation of Comet 1P/Halley was made on 9 February 1910 UT, also by Maximilian Wolf in Germany (Kronk, 2007: 147; Seargent, 2009: 55). Tenn (1994: 27–28) points out that Wolf "… won the race with Barnard and others to be the first to recover Comet Halley." The comet reached perihelion on 20 April 1910 UT at a distance of 0.58721 au (88 million kilometres) from the Sun (NASA, JPL). Comet 1P/Halley brightened until it was seen in the daytime sky without optical aid by several observers around the world, including J.B. Bullock of Hobart, Tasmania (Australia), at a latitude of 42° South and longitude 142° East, who blocked out the Sun with his house and saw the comet with the naked eye on 19.01 May (UT) (Seargent, 2009: 57).

After perihelion on 20 April (UT), as Comet 1P/Halley receded from the Earth and Sun, it faded. It was last detected on 16 June 1911 (UT) on a 40-minute photograph taken by Heber Doust Curtis (1872–1942; Aitken, 1943) at Lick Observatory with the 36-inch (91-cm) Crossley Reflector when the comet was at a distance of 5.44 au (816 million kilometres) from the Sun (Kronk, 2007: 140, 160). 1P/Halley approached Earth from below (south of) the ecliptic and passed through the ecliptic plane on ~7 December 1909 (UT). After reaching its most northerly point as viewed from the Earth, it headed south and passed through the ecliptic again on ~4 April 1910 (UT). 1P/Halley was mostly at a low northern declination throughout its naked eye phase (until early June 1910). Figures 3a and 3b compare the orbits of Comets 1P/Halley and C/1910 A1. The Earth's orbit is red, and the comets' trajectories are black.

Now that we have introduced Comet C/1910 A1, we shall probe deeper into the morphological changes of this comet.

## 4 DISCOVERY OF COMET C/1910 A1 AND MORPHOLOGICAL CHANGES

In late 1909 and early 1910, anticipation of Comet 1P/Halley's 1910 apparition was gaining momentum. The excitement about the approach of the comet increased from September 1909, when Wolf made the first photographic observations of the incoming comet, confirming its approach. This is evident from the increase in the number of published New Zealand newspaper articles that followed that detection (see Figure 1). The appearance of a bright, naked eye comet (C/1910 A1) in January 1910 caused confusion for those not abreast with the literature on Halley's Comet, with many misidentifying the bright comet as Comet 1P/Halley.

On 16 January 1910 (UT), Robert T.A. Innes (1861–1933; Orchiston, 2001; 2003) carried out three observations of Comet C/1910 A1 at one-hour intervals from Transvaal Observatory in South Africa (Innes, 1910: 11). Based on those observations, and extrapolating the comet's motion back to the morning of 13 January (UT) using GUIDE 9.1 (2020) planetarium software we found that C/1910 A1 was near ε Sagittarii (at approximately R.A. 18h 25m and Dec. –34° 30′ for an epoch of 1910) at that time. This agrees with the position of the star Altair. Innes had used this star to test the accuracy of the setting circles on the mounting of the 9-inch refracting telescope before using these circles to record the right ascension of the comet.

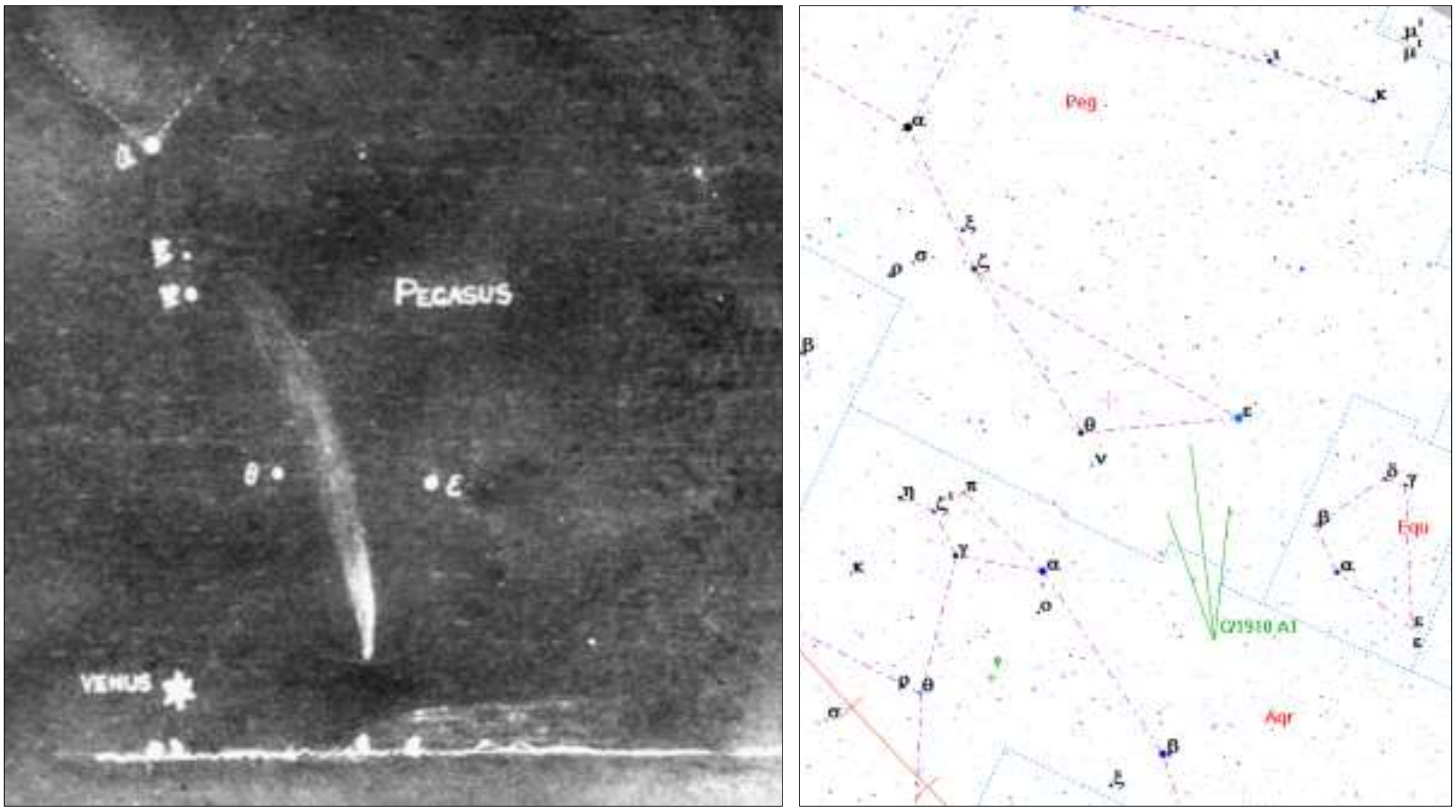

Figure 4: Sketch of Comet C/1910 A1 on 29 January 1910 (UT) displaying a 20° tail (after Rolston, 1910: 409–410). A screenshot from GUIDE 9.1 (2020) Planetarium software is placed alongside for star, comet (in green) and Venus (small green cross) comparison. Note that the comet sketch (left) equates to a tail length of 21° when measured in GUIDE 9.1 (2020). Also note the brightness of the head and tail in relation to Venus (magnitude –4.6) and α Pegasi (magnitude 2.5). The text annotated on the sketch is part of the original illustration.

At discovery, the comet was 11° south-west of the Sun (GUIDE 9.1, 2020) and, thus, a morning object and it was five days from its perihelion encounter with the Sun on 17 January 1910. The Sun–comet distance was reasonably close, ~19 million kilometres (about 0.13 au). C/1910 A1's extreme brightness was a direct result of its close proximity to the Sun and the acute comet–Sun angle as seen from Earth. When they are far from the Sun, in the cold depths of space, comets are dormant. As they approach the Sun their surfaces are heated, and eventually the ices on and near the surface sublime, turning directly from a solid to a gas in the vacuum of space (Haack, 2021). The gas produced in this way erupts from the comet's surface through active areas in vast jets, carrying with it significant quantities of dust and debris (Miles, 2016: 71–72). As a result, the comet's nucleus is rapidly shrouded in a cloud of gas and dust known as the coma, from the Greek word 'kome' meaning 'hairy' (Guillemin, 1877: 197). That gas and dust is then blown outwards by the combined influence of the solar wind and interaction with the Sun's radiation, causing the comet to grow its tails (both dust and gas), which point away from the Sun (Liu et al., 2023). The ancient Roman historian Appian of Alexandria (ca.95–165 CE) is believed to have been the first to recognise that comet tails always point away from the Sun (Liu et al., 2023: 206). Comet C/1910 A1 displayed a long bright tail and a sketch by Rolston (29 January 1910) of this tail is given in Figure 4.

The closer a comet gets to the Sun, the hotter its surface, and the greater the rate at which it will eject dust and gas. As the frozen nuclear ices sublimate with increased volatileness, locked dust is released in the expanding gases (Wyckoff, 1982: 16). As a result, more material is dispersed into space to reflect sunlight to the Earth. In addition, the closer a comet is to the Sun, the more intense the light that the comet will reflect. The combination of these two effects causes comets to rapidly brighten as they approach the Sun—and the closer a comet comes to our star, the brighter it will become. Finally, if a comet passes directly between the Earth and the Sun, or within a few degrees, a process known as 'forward scattering' can act to further increase the comet's brightness, leading to a brightening by up to a factor of one hundred (Marcus, 2007: 119–130).[3]

The rapid brightening and daytime visibility of Comet C/1910 A1 seem very similar to the observed behaviour of the Kreutz Sungrazing comets—which include amongst their number the majority of the brightest daytime comets observed over the past two thousand years (Marsden, 1967). Despite C/1910 A1's relatively small perihelion distance (0.13 au), it was more distant from the Sun than the typical perihelion passage of Kreutz sungrazers, whose perihelia are often within two Solar radii (i.e., within 1.4 million km of the Sun's surface; Sekanina, 2024: 1). We discuss New Zealand observations of Kreutz family comets in detail in a companion paper that is now in preparation (but see, also, Orchiston et al., 2020a).

According to international observations, for four days Comet C/1910 A1 was visible to the naked eye in the daytime. Figure 5 displays a sketch of the comet in relation to the Sun on 17 January 1910 (UT). Observations in Kronk (2007: 171–172) suggest that it was brilliant between 17 and 20 January UT during its perihelion/perigee passage (Aitken, 1910: 29–30; Boss: 1910: 70; Pickering, 1910: 1; Rolston, 1910: 372; The New Comet (1910a), 1910: 440–441), then it slowly faded over the ensuing weeks and was no longer a naked eye comet from mid-February 1910 (Kronk, 2007: 175–176; The New Comet (1910a), 1910: 440). The reason for the rapid decline in magnitude was the comet's growing distance from both the Sun and the Earth (*ibid.*), and reduced forward scattering of light. Table 2 provides an overview of Comet C/1910 A1's evolution based on Kronk (2007: 170–179). It lists the observation date (in UT), the location of the observer(s), the declination of the comet, the brightness (magnitude) of the comet, if the tail was visible and what length it was, in degrees, and key notes relevant to the observation. For information regarding the observers, locations, observation details, and other data, we direct interested readers to Kronk's *Cometography, Volume 3: 1900–1932* (2007).

Using spectroscopy, astronomers noted that C/1910 A1 had a higher ratio of dust than most comets (e.g. see Eicher, 2013: 145). This greater-than-normal dust ratio and close proximity to the Sun greatly enhanced the forward scattering of light effect. According to Seargent (2009: 145), F. Baldet (1885–1964) used a "... small dispersion spectrogram … [which revealed a] purely continuous spectrum of reflected sunlight …" from the nucleus and the tail out to 8°. Seargent (*ibid.*) also states that the dusty nature of the comet was confirmed by Lick Observatory astronomer William Hammond Wright (1871–1959; Shane, 1979). Meanwhile, a five-minute exposure taken from Meudon, France, on 22 January 1910 with red-sensitive photographic plates revealed "... a brilliant nucleus giving a continuous spectrum from Å700 to Å420 and several condensations." (The New Comet (1910a), 1910: 442). Usual cometary emissions were not visible, probably due to the forward-scattering of sunlight overwhelming the weaker emission lines (Seargent, 2009: 146).

A daytime observation on 18 January 1910 (UT) by Wright (Lick Observatory, USA) showed "... the spectrum of the comet's nucleus to be continuous, with the sodium D lines bright." (Pickering, 1910: 1; c.f. Rolston, 1910: 372). Thus, a strong sodium presence was detected (Boss, 1910: 70; Wright et al, 1910: 179–181). This contributed to a "... distinctive yellowish tint …" as seen by visual observers (Brown, 1973: 85). Whittaker even described the head of the comet as "... dusky red …" (Rolston, 1910: 372). The Meudon observers also noted that the brightest spectral condensation was at sodium (590 nm) (The New Comet (1910a), 1910: 442). H.F. Newall (1857–1944; Royal Society, 2025: 716–732) at Cambridge Observatory (England) also observed a strong sodium presence. Using a direct-vision prism he found that "... sodium was traceable in the side trains to a distance of 7¾′ (minutes of arc) behind the nucleus …" (Newall, 1910: 459–460). Newall's observations as shown here in Figure 6 show that he observed a dark lane extending behind the nucleus, with brighter "... side trains …" on either side. A'Hearn

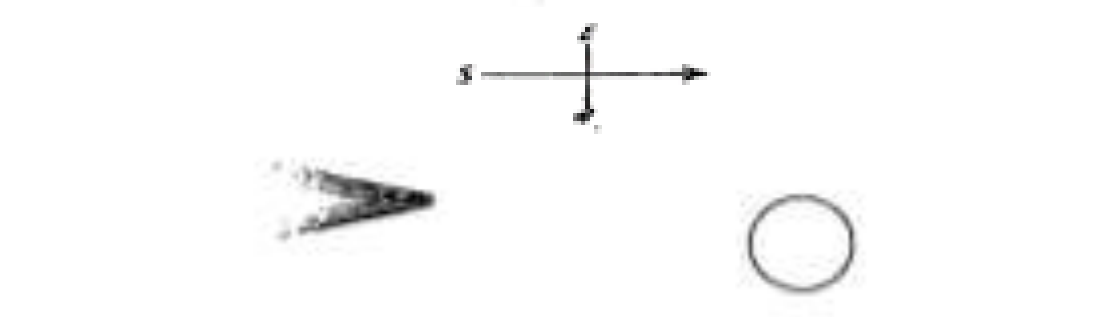

Figure 5: Sketch of Comet C/1910 A1 at perihelion from Johannesburg, South Africa, on Monday 17 January 1910, 2 pm local time (16 hr UT). The circle is the Sun and equates to an apparent diameter of 0.5°, thus, the comet is about 1.5° south of the Sun in this sketch (after Innes, 1913: 11–12).

(1982: 433–434) reminds us that photographic spectroscopy is the classical method for studying spectra. However, in Newall's case, high moisture content in the air and on the telescope lens prevented spectrographic detection, so he had to use the direct-vision prism.

Rolston (1910: 373) informed his readers that similar lines were seen in the tail of Comet Wells (C/1882 F1) and the Great Comet of 1882 (C/1882 R1). E.C. Slipher also detected sodium lines in Comet 1P/Halley with the slit spectrograph at Lowell Observatory, later in 1910 (Lemon and Bobrovnikoff, 1926).

A further observational feature that seems related to the excessive dustiness of Comet C/1910 A1 is that the Type II tail produced a number of synchronic bands, striae and syndynes. These bands were visible in the comet's dust tail and caused the comet to display linear regions of different densities (Steckloff and Jacobson, 2015). Building on the work of Fedor Bredikhim (1831–1904), Sekanina (1976: 389) suggested that "... great spurts of various sized dust particles …" leave the nucleus and then burst into smaller fragments upon encountering intense solar radiation, thereby resulting in mini comets (Sekanina and Farrell, 1980: 1538). Dust particles with diameters of ~1–100 μm are believed to be released from active regions on a comet's surface. Solar radiation pressure causes these particles to escape from the comet to form the banding in the tail (Steckloff and Jacobson, 2015: 1–3, based on the work of Kharchuk and Korsun, 2010: 80–86). Sekanina (1976: 389) emphasises that this phenomenon has "... been exhibited by only a handful of comets." He notes (*ibid.*) that the best known of these were C/1910 A1, as well as Donati 1858 VI (C/1858 L1), Mrkos 1957 V (C/1957 P1), and Seki-Lines 1962 III (C/1962 C1). Kharchuk and Korsun (2010) add C/2006 P1 (McNaught) to this list (see, also, Price et al., 2019). Figure 7 shows a photograph of C/1910 A1 taken on 29 January 1910 (UT) from Lowell Observatory, USA (left; Hale, 2020) with the striae lines visible in the dust tail. A photograph of C/2006 P1 (McNaught), is placed alongside, for comparison, where C/2006 P1 also displayed beautiful striae in its dust tail. Note, also, the comparative sketch by A.B. Douglas of the growing length of the tail of C/1910 A1 in Figure 8.

Table 2: Observational overview of the naked eye apparition of the Great January Comet of 1910 (C/1910 A1), with dates listed in Universal Time (based on Kronk, 2007: 170–175). The magnitude and tail lengths are sometimes averaged across the range of observational estimates on a given day, as indicated by red font. Countries where the observations are from are listed. Note that several observations were made from the same country simultaneously. Observations from New Zealand are in green print and are based on New Zealand newspaper articles—not Kronk or Vsekhsvyatskii. 'Dec' is the comet's declination (based on GUIDE 9.1 (2020) planetarium software). 'Magnitude' is the comet's apparent magnitude, which also is based on GUIDE 9.1 (2020).[4] 'Tail' is the comet's tail length in degrees. In the 'Notes and Comments' column, 'Naked eye' denotes naked eye observations for which no apparent magnitude is recorded, while '?' means the magnitude is uncertain.

| UT Date | Location (Country) | Dec (°) | Magni-tude | Tail (°) | Notes and Comments |
|---|---|---|---|---|---|
| January 1910 | | | | | |
| 13 Jan | South Africa | –30 | –1 | Yes | 'Discovered' from Cullinan, South Africa, by diamond miners. Naked eye. |
| 14 Jan | South Africa | –30 | –1 | Yes | Seen again by the diamond mine workers. Naked eye. |
| " | New Zealand | –30 | NE | Yes | Seen from Gisborne and near Queenstown, NZ (15 Jan NZST). Naked eye. |
| 15 Jan | —— | –29 | | | |
| 16 Jan | —— | –28 | | | |
| 17 Jan | South Africa | –27 | –5 | 01 | Daytime comet! Perihelion. Brighter than magnitude –5? |
| 18 Jan | Alger (Algeria), Austria, India, Italy, New Zealand, South Africa, USA. | –24 | –5 | | Daytime comet! Seen from Mount Egmont, NZ (dusk?). Brighter than magnitude –5? |
| 19 Jan | Alger (Algeria), Chile, England, India. | –21 | –5 | 02 | Daytime comet! Two tails. Brighter than magnitude –5? |
| 20 Jan | Alger (Algeria), England, Germany, India. | –17 | –4 | 05 | Daytime comet! Fading. |
| 21 Jan | Belgium, Germany, Netherlands, Poland. | –14 | –3 | 07 | No longer a daytime comet. Bright evening comet. |
| 22 Jan | Belgium, France, Germany, New Zealand, Poland, USA. | –11 | 1 | 06 | Seen from the eastern slopes of Mount Egmont, NZ (unsure of exact date). |
| 23 Jan | Germany, Italy, Spain, Sweden. | –09 | 1–2 | 15 | 15° tail visible to the naked eye. |
| 24 Jan | Alger (Algeria), Poland. | –07 | | 15 | Tail bent towards the east. |
| 25 Jan | France, Spain, USA. | –05 | 1 | 10 | |
| 26 Jan | Alger (Algeria), Belgium, France, Hungary, New Zealand(?). | –03 | | 22 | Remarkably reddish head (through the telescope). Seen (?) from the Waikato region (near Hamilton), NZ. |
| 27 Jan | Belgium, Egypt, France, Germany, Italy, Poland, Netherlands, USA. | –02 | 2 | 25 | Forked tail. |
| 28 Jan | Germany, Italy, Sweden, USA. | –01 | 2 | 50 | Nucleus had a north–south axis on photographs from Lick Observatory. |
| 29 Jan | Belgium, France, Germany, Russia, USA. | 00 | 2.7 | 30 | |
| 30 Jan | France, Germany, Russia, USA. | +01 | 3.5 | 27 | Fainter than the previous evening (Baranow). |
| 31 Jan | Belgium, France, Germany, India, Italy, USA. | +02 | 3 | 35 | Tail merged with the zodiacal light (van Biesbroeck). |
| February 1910 | | | | | |
| 01 Feb | Belgium, France, Netherlands. | +02 | 3.5 | 30 | Tail very long and thin to the naked eye (Chofardet). |
| 02 Feb | France, USA, Netherlands. | +03 | 4.8 | 16 | |
| 03 Feb | —— | +04 | | | |
| 04 Feb | —— | +04 | | | |
| 05 Feb | Alger (Algeria), France, Switzerland, USA. | +05 | 5 | 25 | Comet 'enormously weakened' (Guillaume). |
| 06 Feb | —— | +06 | | | |
| 07 Feb | Alger (Algeria), Italy. | +06 | 4.5 | 20 | 20° tail seen with the naked eye. |
| 08 Feb | —— | +06 | | | |
| 09 Feb | Belgium, Germany, USA. | +07 | 5.5 | | Very faint trace of a tail (Douglass, USA). |
| 10 Feb | France, Germany, Russia, Switzerland. | +07 | 6–7 | | |
| 11 Feb | —— | +08 | | | |
| 12 Feb | Belgium, Russia. | +08 | 6.5 | | Now too faint to be seen with the naked eye. |

Numerous observations of the changes in Comet C/1910 A1's brightness, tail length, outpouring of dust and sodium, and features of the coma were observed by international astronomers. What of New Zealand astronomers? Did they, or the public, contribute to our understanding of this remarkable comet?

## 5 OBSERVATIONS OF COMET C/1910 A1 FROM NEW ZEALAND COMPARED TO INTERNATIONAL OBSERVATIONS

Based on newspaper reports, relatively few observations of C/1910 A1 were made from New Zealand. The first reported sighting was from Gisborne on 15 January 1910 NZST (14 January UT), one day after initial overseas observations (14 January NZST, 13 January UT). The *Gisborne Times* states "A comet, presumably Halley's Comet, was visible in the east just above the horizon, about sunrise this morning." (*Gisborne Times*, 1910). No other New Zealand papers reported this observation. No direct magnitude comparison with other astronomical bodies is given, no tail is mentioned, and nor is the observer named. To be visible at sunrise is a clear indication of the brilliance of the comet. Gisborne is on the east coast of the North Island of New Zealand, so the observer would most likely have had an exceptionally low ocean horizon, particularly if they were observing from (for example) Wainui Beach. Once again, the author of the newspaper article displayed the common mistake which many succumbed to in relation to Comet C/1910 A1, "A comet, *presumably Halley's comet* was visible …" To the general public, Halley's Comet was coming, therefore if a bright comet appeared in the east before sunrise it *must* be Halley's Comet. The more astronomically astute would have correctly differentiated between the two. For the date in question, GUIDE 9.1 (2020) planetarium software indicates that C/1910 A1 was rising at 4:00 am local time and was located in the constellation of Sagittarius. Sunrise for Gisborne on that date was 4:59am (NZST). The comet had a western elongation of 11° from the Sun (*ibid.*), so rose in late astronomical/early nautical twilight. In regards to magnitude, the Gisborne observation of the comet was presumably similar to that of the South African miners the previous morning, but may have been brighter as C/1910 A1 progressed towards perihelion on 17 January 1910 (UT), suggesting that the comet could have been as bright as apparent magnitude –1 or even –1.5.

After the sighting of Comet C/1910 A1 from Gisborne was published in the *Gisborne Times* (1910), one might expect that at least some of the readers of that newspaper would want to see the comet for themselves. This does not appear to be the case. There are surprisingly few other accounts of sightings of C/1910 A1 from New Zealand. Indeed, no other newspapers reported this 15 January observation from Gisborne in their own publications. It should be noted that mid-January is mid-summer in New Zealand, usually characterised by hot, sunny weather, so theoretically, many clear nights.

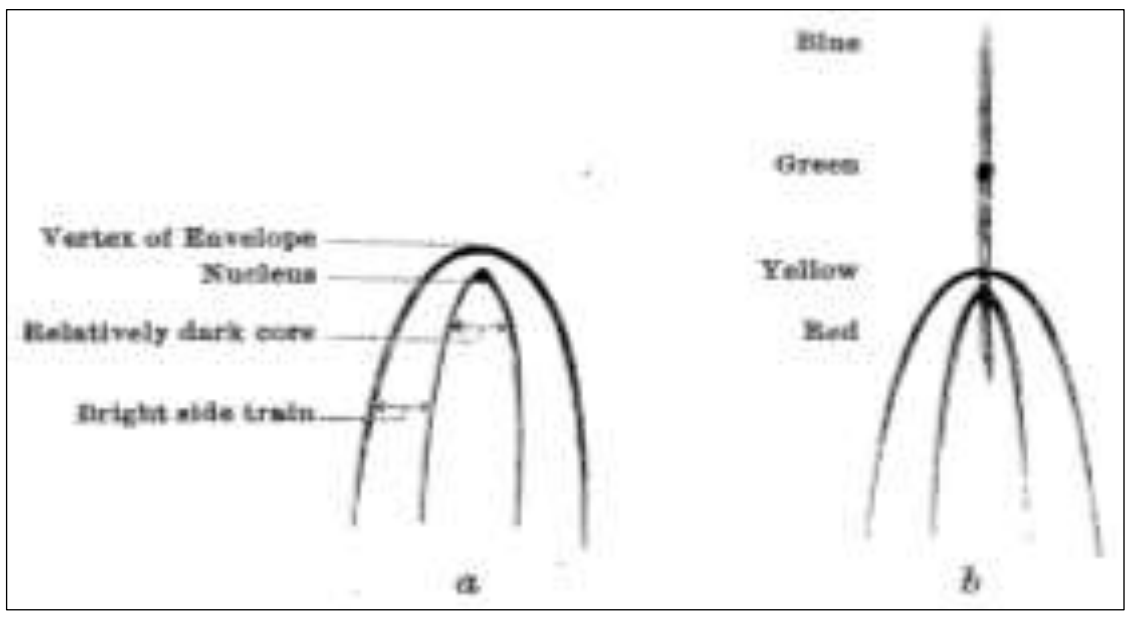

Figure 6: Spectral sketches of Comet C/1910 A1 as viewed visually through a direct-vision prism with 25-inch and 6.5-inch refractors at Cambridge University's Observatory by H.F. Newell on 22 January 1910. Upon inspection of Newall's sketches and observations, British astronomer A.R. Hinks (1873–1945) concluded that the "... observations showed that the spectrum of the comet's tail was purely monochromatic, the one line being due to sodium or helium …" (after The New Comet (1910a)).

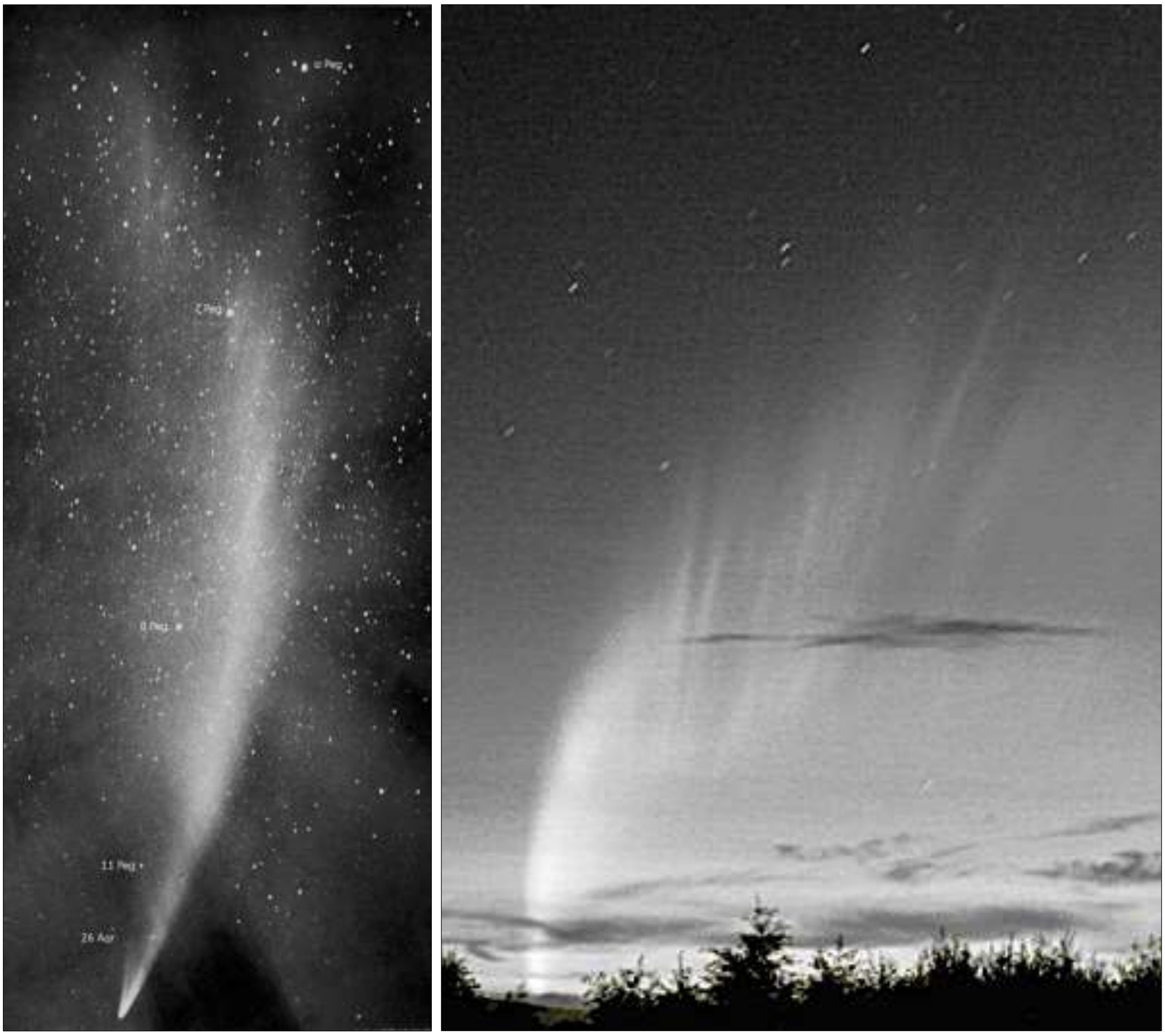

Figure 7: A comparison between the tails of Comets C/1910 A1 (left) (Earthrise, n.d.) and C/2006 P1 (McNaught) (right) (Drummond, 2007). Note the synchronic bands in the tails of both comets. C/1910 A1 was photographed from Lowell Observatory (Arizona, USA) by Carl Lampland and Vesto Slipher on 28 January 1910 (UT), 11 days after perihelion (Bortle, 2010). C/2006 P1 (McNaught) was imaged by the first author of this paper (John Drummond) on 19 January 2007 (NZDT) from New Zealand. The image of Comet McNaught has been adjusted to mimic the appearance of a photograph taken with the technology available to observers in 1910, to provide a comparison of the striations between the two comets. It should be noted that, despite their use to compare the structure of the two comet's tails, the images are not identical in scale/size.

Ten days after that first reported observation (25 January 1910 NZST), an article in the Queenstown *Southern Standard* newspaper recounted the experience of Mr T.P. Laffey, who with others ascended Ben Lomond Peak on 15 January (NZST) and saw a comet "... low down near the horizon." (*Lake Wakatip Mail*, 1910). Once again, the observed object was assumed to be Halley's Comet. The altitude of Ben Lomond Peak is 1,748 metres (Ben Lomond, n.d.), and it is located five kilometres NW of Queenstown (see Figure 9). Since Ben Lomond Peak lies approximately 9.4° west of Gisborne, both comet-rise and sunrise would have been later on the mountain than on the beach at Gisborne. As such, the observations from Gisborne would have been made around 40 minutes earlier than those from Ben Lomond, despite occurring on the same morning. The tail of the comet as seen from Ben Lomond Peak was described as rising vertically, and broadening towards the top. After the daytime appearance (17–20 January UT), when the comet moved away along a subtended angle from the Sun, Charles Michie Smith (1854–1922; Kamersara Rao et al., 2014), the Government Astronomer at Madras Observatory in India (Kapoor, 2023: 411), also described a similar tail appearance several days later (26 January UT). He wrote that the tail was 22° long, and he said that it was fairly straight until near the end "... when it bent away southwards, and mingled with the zodiacal light." (Kronk, 2007: 173).

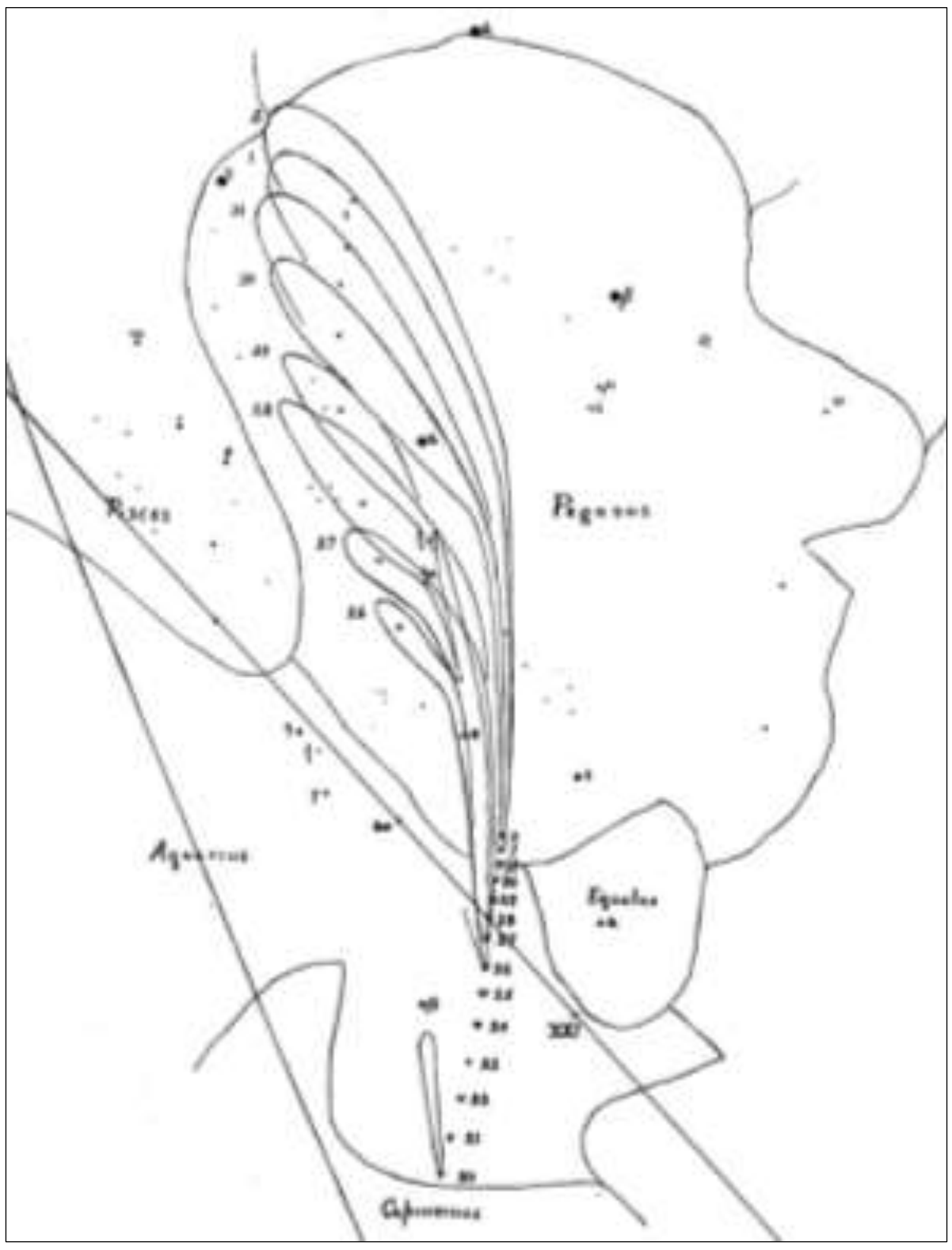

Figure 8: Drawings of Comet C/1910 A1 and its impressive tail as drawn by A.B. Douglas. Note the development of the tail from 20 January until 2 February 1910. The stars of the Square of Pegasus are evident. The authors of this research paper measured a diagonal line through the Square of Pegasus as 21°, thus, the immense length of the tail is apparent. The curved lines are constellation boundaries (after Douglas, 1910: 162).

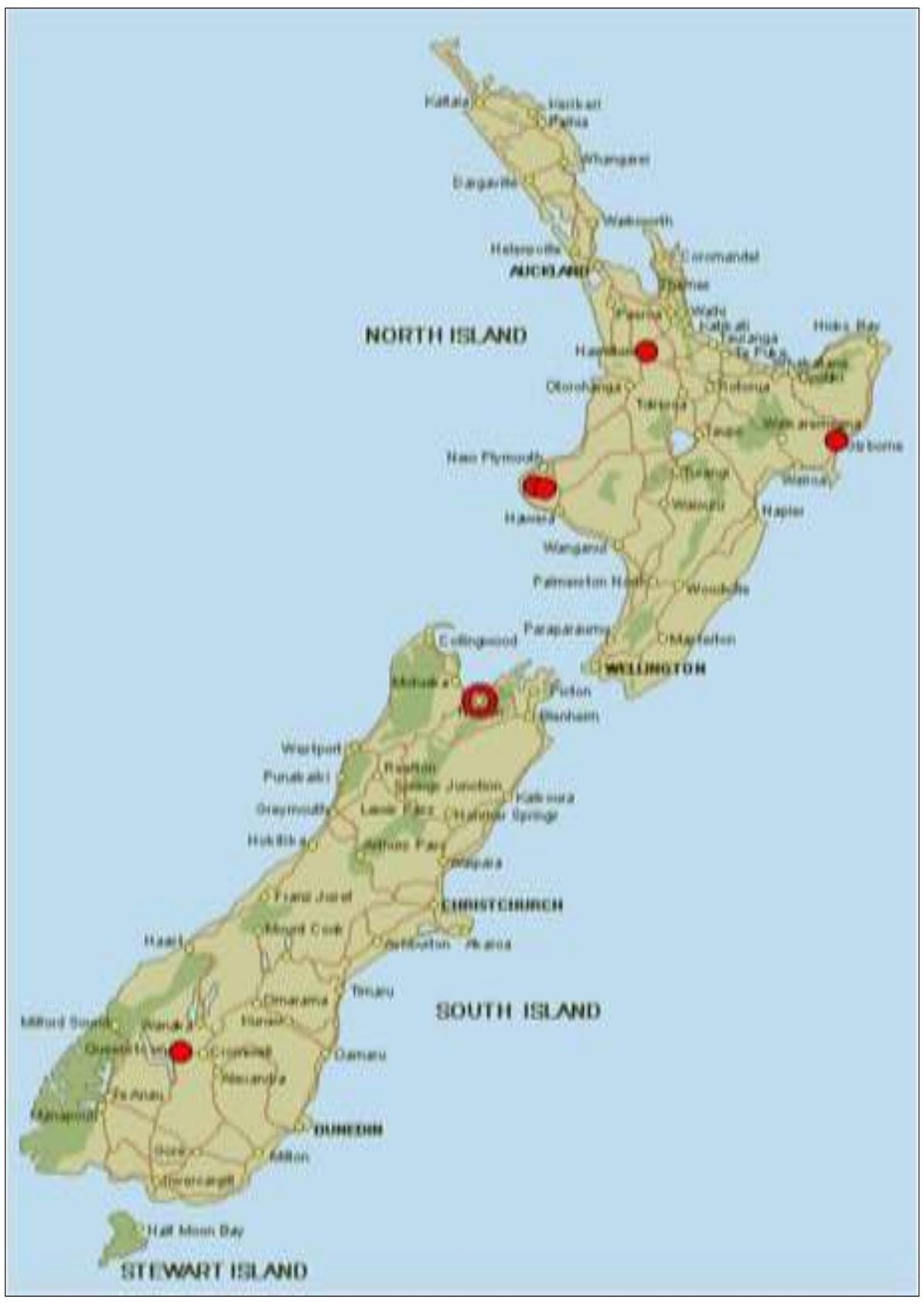

Figure 9: Map of New Zealand showing the locations from which Comet C/1910 A1 was observed in 1910, based on our results obtained from the Papers Past archive; the open circle marks a non-newspaper report by Dennis Goodman's grandmother (base image: Backpack New Zealand; map modifications: the authors).

At perihelion, when Comet C/1910 A1 was brightest, no observers from New Zealand are reported to have seen it in the daytime sky. W.E. Rolston (1876–1921; H.F.N., 1922) reiterated the contemporary name for it as the Daylight Comet in *Nature* (Rolston, 1910: 372). He received reports that W.M. Worssell and R.T.A. Innes from Transvaal Observatory in Johannesburg observed the comet on 17 January at 9:29 am local time (when the Sun was well above the horizon) (*ibid.*). Innes observed it with the naked eye at midday, and stated that it had a 1° fan-shaped tail when 4.5° west of the Sun (*ibid.*). Observers at Lick Observatory (in the USA) stated that "... it was several times *as bright as Venus* at its maximum brilliance …" (Wright et al., 1910: 179; our italics), so around magnitude –6 to –8. It was described as having a head 5′ in diameter and a well-developed 1° fan-shaped tail, visible to the naked eye. Rolston (1910: 372) also reported that observers at Milan Observatory (Italy) saw the comet "... in full daylight." Rolston (*ibid.*), Kronk (2007: 171–172) and Sergeant (2009: 144–145) also wrote of daytime observations from Algeria, Austria, Chile, England, India, Ireland, Italy, Scotland, South Africa, and the USA (see Table 2). Between them, the authors of this paper have witnessed one daytime comet, namely C/2006 P1 (McNaught) in January 2007. It was an exceptionally surreal experience. A photograph of C/2006 P1 taken in broad daylight by the first author of this paper (JD) is shown in Figure 10 to illustrate how C/1910 A1 likely appeared to observers in 1910.

From 19 January 1910 (NZST), newspapers across New Zealand reported that a bright comet was seen several days earlier from Johannesburg, South Africa (A Comet, 1910; The

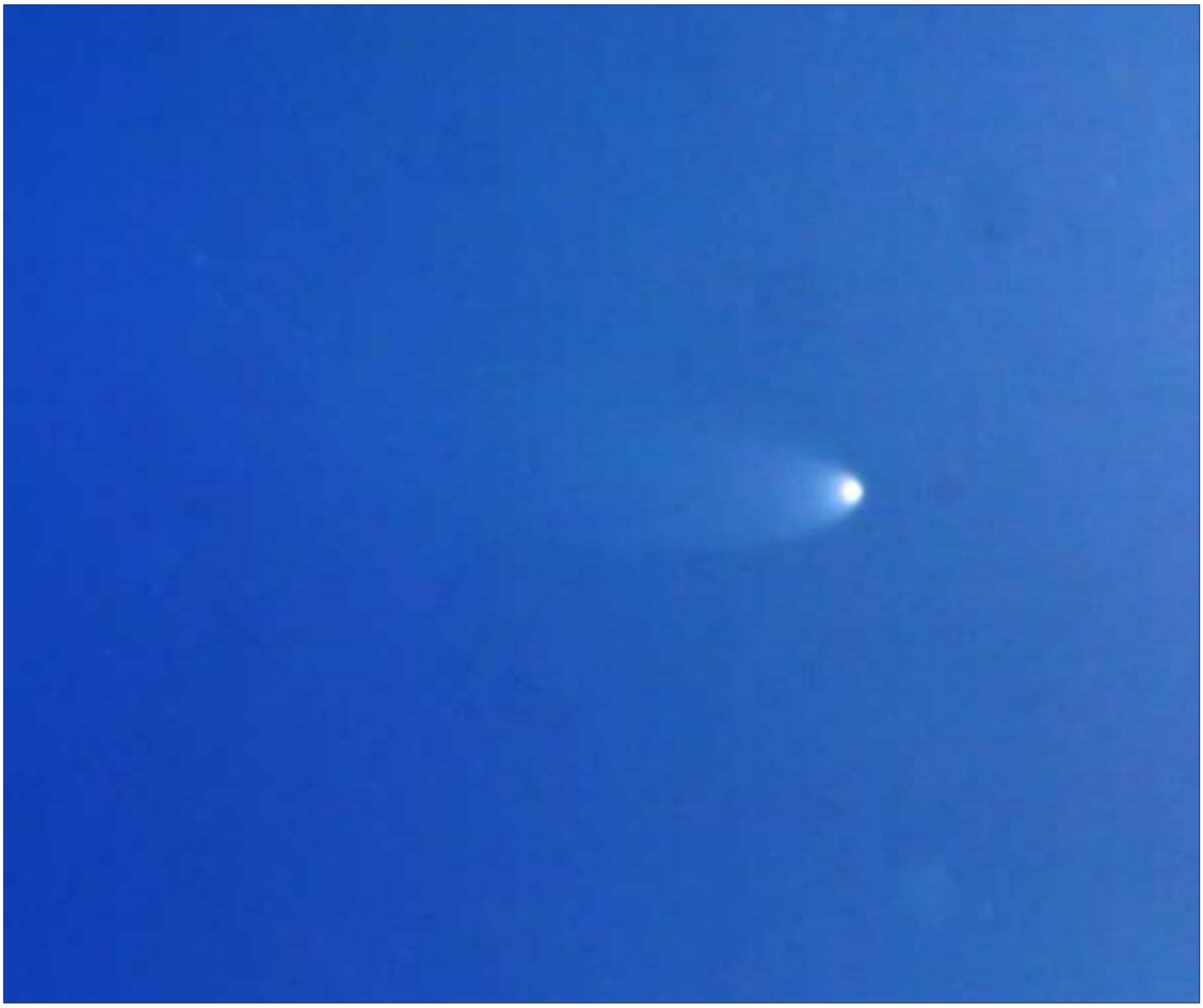

Figure 10: Comet McNaught (C/2006 P1) photographed by the first author of this paper during the daytime on 14 January 2007 (NZDT). The image was taken using an 80-mm f/7.5 refractor and Canon DSLR camera. Comet McNaught was visible to the naked eye in the daytime for one day, and was reported to have been the brightest comet in more than five decades. By comparison, Comet C/1910 A1 was a naked eye daytime comet for four days.

Trail of a Comet, 1910; Was it Halley's Comet? 1910). The comet was described as being 5° south-south-west of the Sun at sunrise and still visible after sunrise. It was stated to be moving 4° eastward and 2.5° northward each day (at that stage) (A Comet, 1910). Around the same date, the *Otago Daily Times* published an article called "The Night Sky" in January (20 January 1910) in which the positions of the constellations, planets, and Comet 1P/Halley are mentioned. There was no word about this new, bright comet in the January sky. Perhaps the article in question was written before the one that described the new comet on 19 January (NZST), the day before. According to Papers Past, nearly 30 articles were published in the 21 or 22 January newspapers about observations of the January comet being made from England on ~20 January 1910 (The Latest, 1910).

In the 25 January edition of the *Wanganui Herald*, Joseph Ward (1862–1927; Orchiston, 2016: 301–335), the Director of the Wanganui Observatory in New Zealand, wrote that he received word that a resident from Taranaki saw Comet C/1910 A1. The observer ascended Mount Egmont (now Mount Taranaki) on 19 January (NZST) and saw the comet "... before the cabled report from Johannesburg was received …" (Local and General, 1910a). The summit of this mountain is 2,518 metres above sea level. This observation was reported in eight newspapers around the country (Notes from Wanganui, 1910; The Comet, 1910; Local and General, 1910b). Perihelion occurred on 17 January UT, two days prior to the ascent, and the comet had moved to an eastern elongation (and therefore was visible in the evening sky). On 19 January (NZST) it was approximately 6° east of the Sun, and set about ten minutes after sunset. Modelling using the GUIDE 9.1 (2020) planetarium software indicates that the comet was around magnitude –4.7 at that stage. Perhaps the elevated altitude of Mount Egmont afforded excellent clarity to view Comet C/1910 A1. The Taranaki observer undoubtedly saw the comet during this brilliant post-perihelion stage, probably at early sunset. It is entirely possible that they also saw the comet in the daytime sky before the Sun had set; however, none of the newspaper articles states this.

Also, from the Taranaki region, a group of half a dozen people were reported to have seen Comet C/1910 A1 from Dawson Falls, on the eastern slope of Mount Taranaki, at an altitude of 902 metres (Dawson Falls Area, n.d.). This was published in the *Taranaki Daily News* (Local and General, 1910a) which quoted the *Eltham Argus*, a Taranaki newspaper (1897–1967), which unfortunately is not covered by Papers Past. No definite date was given for this observation, but we estimate that it may have been around 18–21 January. According to GUIDE 9.1 (2020), from 22 January 1910 the comet was setting within a few minutes of the Sun and lay about 15° north-east of it, along a line parallel to the horizon.

On 27 January 1910 NZST, a correspondent wrote to the *Waikato Independent* (1910) newspaper (based in Hamilton) stating that

> Halley's Comet was visible last night at about 8.5 o'clock [i.e. 8.30 p.m.]. Its tail reached from north-east to east, covering about 90 degrees. The rising of the moon spoilt the effect, as the comet is somewhat faint as yet.

On that night (26 January NZST), GUIDE 9.1 (2020) software (set for Hamilton's location) indicates that the comet's head set at about 7:32 pm local time, and the Sun set a few minutes later. The Moon was, at that time, 14.7 days old, just past Full Moon, and rose at about 8:28 pm local time, an hour after the comet had set and towards the end of nautical twilight. Note that the head, having set an hour earlier, was 12° below the west-south-western horizon and yet they saw a tail "... covering about 90 degrees." This (~90° + 12°) equates to a tail length of approximately 100°.

Observers mentioned in Kronk (2007: 173) for ~26 January 1910 (UT) state that the tail was only about 25° long. On 29 January (NZST) overseas observers were reporting a 50° tail (*ibid.*). This therefore raises the question as to whether this New Zealand observer in the Waikato actually saw Comet C/1910 A1's tail as ~100° in length. The observer described the tail as reaching "... from north-east to east." This could have meant that the end of the tail was coming from or towards the north-east. One caveat is that GUIDE 9.1 (2020) software suggests that the comet's tail was oriented approximately north along the horizon, however observers around this date noted that the tail bent towards the east. Some, such as C.M. Smith in India, said the straight tail "... bent away southwards…" (*ibid.*), so perhaps some of it did extend above the western horizon in the evening dusk sky.

Alternatively, perhaps the observer saw the comet's tail *and* the zodiacal light, a phenomenon that some in the Northern Hemisphere said interfered with viewing the comet's tail. Indeed, Miss Eleonora Armitage from Herefordshire in England stated that the glow that some saw in the neighbourhood of the comet was the zodiacal light. She recalled that it was very strong on 29 January, with both the comet's tail and zodiacal light having the "... same degree of luminosity." (The New Comet (1910a): 440). Robert G. Aitkin (1864–1951; van den Bos, 1958) agreed, stating the comet's 30° tail "... was lost in the zodiacal light …" (Aitken, 1910: 30). Several other observers also noticed the apparent entanglement of the comet's tail with the zodiacal light at the end of January (*ibid.*). So, in reality, the Waikato observer possibly saw the tail of the comet but more likely observed the zodiacal light. In the evening skies in January from New Zealand, the ecliptic, and thus the zodiacal light, runs at an angle of approximately 45° to the horizon, heading towards the north-east.

By the end of January 1910, the comet was observed to be fading. In his *Astronomical Notes for February* the Dunedin-based Reverend Paul Wynyard Fairclough (1852–1917) wrote,

> We have not yet heard of the Transvaal Comet being seen in New Zealand, but we know of no reason why it should not be seen, except that the mornings have been cloudy, that there is a bright moon, and that the comet itself is fading rapidly. (Fairclough, 1910).

In his paper, Fairclough quotes from several international sources, and he possibly learnt of the comet fading from them. Fairclough's statement that "... we have not yet heard of the Transvaal Comet being seen in New Zealand …" is likely due to reports not being published in his local newspaper, thus highlighting the lack of dissemination of this comet's observations across New Zealand.

Based on research using Papers Past, asking New Zealand observatories for observations of Comet C/1910 A1, and accessing Carter Observatory files (for a while it was New Zealand's national observatory and repository of astronomical observations), the observations of C/1910 A1 mentioned here are possibly the only ones reported from New Zealand (see Figure 9 for observers' locations). Some observations may have been distributed as letters using the 'one penny universal postage', and other observations not distributed but kept as personal records. An example of this is from the Royal Astronomical Society of New Zealand's Past President, Dennis Goodman (pers. comm., 31 August 2025), who relayed to the first author that his grandmother saw Comet C/1910 A1 from Nelson in January 1910:

> Grandma (Eva Barltrop) didn't have anything in writing. It was all as she remembered as a 17-year-old girl living in Nelson. She told me a few times about the two comets in 1910, and how some people thought the January comet was Halley's Comet. I think Grandma initially thought so too, but when the real Halley's Comet appeared some months later she realised she'd been privileged to see two naked eye comets within just a few months! She told me the January comet was very bright in the morning sky but visible only for a very short period of time. I think she mentioned just days. I can't recall much more detail than that, sorry. For Halley's Comet she certainly recalled how the Earth passed through the tail of the comet and how spectacular the comet appeared. This isn't a lot of information, sorry. Just what I recall as my grandma took note of my developing interest in astronomy.

To date, no other personal or penny post observations of C/1910 A1 have surfaced for inclusion in this paper. Compared to the ensuing published observations of Halley's Comet a few months later, the number for Comet C/1910 A1 is disappointing. British-born Professor Hugh William Segar (1868–1954; Nield, n.d.), a mathematician from Auckland University College, stated in 1910:

> Of the daylight comet in 1910, I can find no reference in existing (local) journals, but Halley's Comet, which followed closely upon its heels, aroused considerable interest. (cited in Walker and Blow, 1985: 22).

Mackrell (1985: 105) agreed, stating, "... judging by the lack of attention paid to it [Comet C/1910 A1] in the New Zealand press, it was hardly noticed here." Current New Zealand comet enthusiast and astronomy historian Ian Cooper (pers. comm., 30 October 2025) also supports this claim, stating that he only has one account of the Great January Comet of 1910 on file, which is one of the newspaper reports mentioned here in our research paper. We had hoped that Papers Past would bring numerous observations to light, however, apart from the few that did surface, we must conclude that few New Zealanders observed and reported Comet C/1910 A1. It should be noted that the comet was only visible from New Zealand between mid- and late-January 1910 (refer to Table 3).[5]

The authors believe that a key opportunity for New Zealand to contribute to international cometary studies was missed with Comet C/1910 A1. Numerous telescopes of modest aperture were scattered around the country (e.g. see Mackrell, 1985; Orchiston, 2016), some with photographic capabilities. For example, had circumstances been different, invaluable photographs and sketches of the comet's morphological changes could have been made from 15 to 22 January 1910 using the 9-inch Cooke photovisual refractor at Meeanee Observatory, near Napier (see Mackrell, 1985; Orchiston, 2022: 19–24). In addition, detailed observations of the head could have been made under high magnification. If an observatory had a spectroscope, spectra could also have been observed, revealing the recently discovered sodium D lines. Daytime observations perhaps could have been made, with important changes in the comet's head being observed from New Zealand's unique longitude and latitude when the comet was not visible from other countries. However, as far as our research has revealed, the following account summarises the only published observations of Comet C/1910 A1 known from New Zealand.

The comet was first seen from Gisborne and then Queenstown on 15 January, just one day after the initial discovery from South Africa. It was also seen from Nelson, but no written account was kept. The brilliance was noted as it shone in the dawn sky, and the bright tail was observed. A few days later, it was seen from at or near the summit of Mount Egmont (now Mount Taranaki) and a little later from its eastern slopes. Lastly, it may have been seen from the Waikato region, although some details of the observation seem doubtful. The locations from which Comet C/1910 A1 was seen, as reported in this paper, are shown in Figure 9.

The Papers Past articles presented in this paper highlight the confusion that the public, both in New Zealand and internationally, had in differentiating between Comet C/1910 A1 and Comet 1P/Halley. On 4 February 1910, the editor of the *West Coast Times* (1910) sought to clarify to his readers the difference between the two:

> As an impression is abroad that the comet recently seen in Johannesburg is identical with Halley's comet, for the information of our readers we might state that such is not the case. The Johann-

Table 3: Visibility of Comet C/1910 A1 as seen from New Zealand. Note how the comet was initially a bright comet in the morning sky and then rapidly moved north (and to a right ascension similar to the Sun) to the point where it was no longer visible from New Zealand. Based on GUIDE 9.1 (2020) planetarium software, set for Gisborne (177° 52′ 59″.92 E, 38° 37′ 47″.64 S), New Zealand. The following explains the abbreviated headings: NZST: New Zealand Standard Time; C/Rise = comet-rise time; C/Set = comet-set time (red font indicates that the head of Comet C/1910 A1 was below the horizon at sunrise or sunset); Elong: degrees of elongation from the Sun, M = morning sky and E = evening sky; Dec: declination of Comet C/1910 A1 at that date; Const: the constellation where the comet was located on that date; Mag: magnitude (brightness) of the comet; Sun (au): distance from Comet C/1910 A1 to the Sun in astronomical units; Earth (au): distance from Comet C/1910 A1 to the Earth in astronomical units; NZ: Comet C/1910 A1 could be seen from New Zealand, morn = seen in the morning sky, even = seen in the evening sky, both = seen in both the morning and evening sky.

| NZST | Sun-rise | C/Rise | Sun-set | C/Set | Elong (°) | Dec (°) | Const | Mag | Sun (au) | Earth (au) | NZ | Comments |
|---|---|---|---|---|---|---|---|---|---|---|---|---|
| 01 Jan | 04:46 | 03:53 | 19:36 | 19:14 | 09° M | –27° | Sgr | 4.0 | 0.62 | 1.57 | morn | Undiscovered. |
| 07 Jan | 04:51 | 03:39 | 19:37 | 19:14 | 12° M | –29° | Sgr | 2.1 | 0.45 | 1.36 | morn | Undiscovered. |
| 13 Jan | 04:57 | 03:41 | 19:36 | 19:25 | 13° M | –30° | Sgr | –1.2 | 0.25 | 1.09 | morn | |
| 14 Jan | 04:58 | 03:46 | 19:35 | 19:26 | 12° M | –30° | Sgr | –2.0 | 0.22 | 1.05 | morn | First seen from SA. |
| 15 Jan | 04:59 | 03:54 | 19:35 | 19:34 | 11° M | –29° | Sgr | –2.9 | 0.19 | 0.99 | morn | Seen from Gisborne and near Queens-town. |
| 16 Jan | 05:00 | 04:07 | 19:35 | 19:39 | 09° M | –28° | Sgr | –3.9 | 0.16 | 0.94 | both | |
| 17 Jan | 05:01 | 04:26 | 19:34 | 19:43 | 06° M | –27° | Sgr | –4.7 | 0.14 | 0.90 | both | |
| 18 Jan | 05:02 | 04:52 | 19:34 | 19:46 | 03° M | –24° | Sgr | –5.0 | 0.13 | 0.87 | both | Perihelion. |
| 19 Jan | 05:03 | 05:20 | 19:33 | 19:46 | 04° E | –20° | Cap | –4.7 | 0.14 | 0.86 | even | Seen from Mount Taranaki. |
| 20 Jan | 05:04 | 05:44 | 19:33 | 19:44 | 07° E | –17° | Cap | –4.0 | 0.16 | 0.87 | even | |
| 21 Jan | 05:05 | 06:04 | 19:32 | 19:41 | 10° E | –14° | Aqr | –3 | 0.19 | 0.89 | even | |
| 22 Jan | 05:07 | 06 19 | 19:32 | 19:37 | 13° E | –11° | Aqr | –2.1 | 0.22 | 0.93 | even | |
| 23 Jan | 05:08 | 06:40 | 19:31 | 19:33 | 16° E | –07° | Aqr | –0.9 | 0.28 | 0.99 | even | Head setting with Sun. |
| 24 Jan | 05:09 | 06:48 | 19:31 | 19:29 | 18° E | –05° | Aqr | –0.3 | 0.31 | 1.03 | Tail | Head not visible from NZ. |
| 25 Jan | 05:10 | 06:54 | 19:30 | 19:25 | 19° E | –04° | Aqr | 0.3 | 0.35 | 1.07 | Tail | Tail possibly seen from eastern Mount Tara-naki; unsure of date. |
| 26 Jan | 05:11 | 06:57 | 19:29 | 19:21 | 20° E | –03° | Aqr | 0.7 | 0.38 | 1.10 | Tail? | Head not visible from NZ. |
| 27 Jan | 05:12 | 06:57 | 19:28 | 19:17 | 21° E | –02° | Aqr | 1–2 | 0.41 | 1.14 | Tail? | Tail possibly seen from the Waikato. |
| 28 Jan | 05:13 | 07:02 | 19:28 | 19:13 | 21° E | –00° | Aqr | 1.7 | 0.44 | 1.12 | Tail? | Not visible from NZ. |
| 29 Jan | 05:15 | 07:04 | 19:27 | 19:09 | 22° E | +01° | Aqr | 2.1 | 0.47 | 1.22 | Tail? | Not visible from NZ. |
| 30 Jan | 05:16 | 07:07 | 19:26 | 19:05 | 22° E | +02° | Aqr | 2.4 | 0.50 | 1.25 | Tail? | Not visible from NZ. |
| 31 Jan | 05:17 | 07:09 | 19:25 | 19:01 | 22° E | +02° | Aqr | 2.8 | 0.53 | 1.29 | Tail? | Not visible from NZ. |
| 01 Feb | 05:18 | 07:11 | 19:24 | 18:58 | 23° E | +03° | Aqr | 3.1 | 0.56 | 1.33 | No | Not visible from NZ. |
| 07 Feb | 05:25 | 07:10 | 19:19 | 18:35 | 23° E | +07° | Peg | 4.7 | 0.72 | 1.52 | No | Not visible from NZ. |
| 14 Feb | 05:33 | 07:01 | 19:11 | 18:09 | 22° E | +09° | Peg | 6.1 | 0.9 | 1.73 | No | Not visible from NZ. |
| 21 Feb | 05:41 | 06:49 | 19:02 | 17:42 | 22° E | +11° | Peg | 7.1 | 1.06 | 1.91 | No | Not visible from NZ. |
| 01 Mar | 05:50 | 06:32 | 18:51 | 17:12 | 22° E | +13° | Peg | 8.0 | 1.23 | 2.09 | No | Not visible from NZ. |

> esburg comet may be looked for in the northern sky any time during February. Halley's comet will not be seen in New Zealand by the naked eye until the end of March.

When this was written (on 4 February), Comet C/1910 A1 was indeed moving deeper into the northern sky and was in Pegasus (GUIDE 9.1, 2020), setting about 30 minutes before the Sun. Table 3 provides an overview of observation dates and the visibility from New Zealand.

## 6 CONCLUDING REMARKS

In this study we have examined New Zealand newspaper reports discussing the 1910 apparition of Comet C/1910 A1 (the Great January Comet), and the excitement regarding the expected apparition of 1P/Halley later that same year. Both objects were bright naked eye comets, which exhibited long tails. This paper focussed on the first of these comets to reach perihelion, C/1910 A1, whose apparition came as a surprise to astronomers, with the comet having been discovered only days before perihelion. Despite numerous international observations of Comet C/1910 A1, very few sightings were reported from New Zealand. Indeed, no observations from New Zealand are included in either Kronk's (2007) *Cometography Volume 3 (1900–1932)* or in Vsekhsvyatskii's (1964). *Physical Characteristics of Comets*.

We only found five observations reported in New Zealand newspapers in early 1910, but some of these observations were repeated in various other newspapers across the country. We expect that other observations of the comet were made and recorded in personal diaries, or were sent from the observers to their friends through the penny post mail system. However, insofar as hard, printed evidence is concerned, there were far fewer New Zealand observers than we anticipated.

Nonetheless, we compared these New Zealand observations with those made internationally and saw that they generally agreed in their descriptions of Comet C/1910 A1. One observation from the Waikato region could have had zodiacal light interference that possibly made the comet's tail appear longer than it really was, with international observations supporting this conclusion. However, it should be noted that from New Zealand's latitude the zodiacal light does not extend vertically at that time of year.

We also identified a clear trend that afflicted both international and New Zealand comet observers of Comet C/1910 A1: later, they often mistook it for Comet 1P/Halley (which was still several months from perihelion). A thorough investigation of New Zealand newspaper articles using the search word 'comet' revealed strong interest by the New Zealand public in Halley's Comet in early 1910. This interest rose sharply after Maximilian Wolf's September 1909 photographic recovery of Comet 1P/Halley, and peaked when the comet was at its brightest, at perihelion, in May 1910. A detailed analysis of New Zealand observations of Comet 1P/Halley will be presented in a companion paper that is currently in preparation. Only a small number of people saw C/1910 A1, the Great January Comet from New Zealand. The time lag between observation and newspaper reports could have been a contributing factor. Undoubtedly, many people did not know that Comet 1910 A1 was in their mid-January sky.

Comet C/1910 A1 was an important comet primarily because of its sheer brilliance, and therefore it has been designated a 'Great Comet'. For a comet to be visible for four days during the daytime is an incredible quirk of nature. This comet released a large amount of dust, and this led to striations in the tail. This most likely contributed to the comet's luminosity near perihelion, due to the dust being involved in forward scattering of sunlight—an effect that greatly enhanced the brightness of the recent Great Comets C/2006 P1 (McNaught) and C/2023 A3 (Tsuchinshan-ATLAS). C/1910 A1 was also one of the first comets to have sodium detected in its tail. However, all of these traits were observed by international astronomers but not by any New Zealand observers. Comet C/1910 A1 really was a key missed opportunity for New Zealand astronomers.

## 7 NOTES

1. Aoteraroa is the Māori name for New Zealand, and the convention within New Zealand is for both names to be used initially, then for either to be acceptable thereafter. For the convenience of our international readers, we hereafter simply use 'New Zealand' when referring to the country.
2. Papers Past can be accessed at this link: https://paperspast.natlib.govt.nz/. One amusing issue associated with using this system was the number of race horses named 'Comet' that were found—as well as scanned misidentifications with such words as 'cornet', 'come', 'corset' and others. These of course were eliminated from the search results.
3. A good recent example of such extreme brightening was seen with Comet C/2023 A3 (Tsuchinshan-ATLAS). Without the effect of forward scattering, that comet would likely only have reached apparent magnitude +2.5 when it passed between the Earth and the Sun. Instead, it brightened markedly, with observers reporting apparent magnitudes of –3.5 or brighter. The light curve plotted on Gideon van Buitenen's comet page (https://astro.vanbuitenen.nl/comet/2023A3; accessed 2 September 2025) shows the effect of forward scattering in green, with the 'unscattered' light curve in red. Many of the brightest 'Great Comets' gained their 'great' status from such forward scattering effects—including the recent comets C/2023 A3 (Tsuchinshan-ATLAS) which reached magnitude –3 (Green, 2024) and C/2024 G3 (ATLAS) which likely reached a similar magnitude (COBS, n.d.), where both these comets exhibited above-average levels of dust (Moreno, 2025: 949–955; NOIRLab, 2025). It is likely that these combined effects are the reason Comet C/1910 A1 brightened so rapidly around the time of its perihelion passage.
4. One of the anonymous referees noted that

   > It might be a good idea to mention the magnitude parameters of the comet, i.e. *H0* = 5.3, *n* = 4.6 according to Orlow (see Kronk), which is a bit brighter than the average bright comet (*H0* = 6.0, *n* = 4) and maybe provide a graphical light curve. I see that Orlow's values fit the given magnitudes given in the table quite well. This in turn also means that the comet might indeed have been brighter than magnitude –5 (as suspected in Table 2) due to forward scattering as the formula does not take it into account.

   We find these comments interesting, but we feel that such an investigation lies outside the scope of the present paper.
5. The other anonymous referee asked whether poor weather may explain the paucity

of New Zealand observations. We think that this is unlikely. A westerly is the prevailing wind for New Zealand and tends to mostly produce clear weather on the east coasts of the North and South Islands and cloudy weather on the west coasts of both islands (Brenstrum, 2001: 15–17). Conversely, an easterly wind mostly results in clear weather on the west coast. A newspaper article reported that in mid-January 1910 Canterbury was experiencing dry conditions (https://paperspast.natlib.govt.nz/newspapers/GIST19100115.2.6?end_date=15-01-1910&items_per_page=10&query=weather&snippet=true&start_date=15-01-1910&title=GISH%2CGSCCG%2CGIST%2CMATAR%2CPBH%2CPBI%2CPBS%2CTAKIT). January is mid-summer in New Zealand, the hottest and usually sunniest time of the year. During the two weeks in the second half of January when Comet C/1910 A1 was visible from New Zealand, there would have been many people living in populated regions across New Zealand who would have had a chance to view the comet (see, also, Figure,NZ, n.d.).

## 8 ACKNOWLEDGMENTS

We are grateful to Ian Cooper (Palmerston North, New Zealand), Dennis Goodman (Nelson, New Zealand), and Dr. David Seargent (The Entrance, New South Wales, Australia) for providing relevant information, and two anonymous referees for their helpful comments. We also wish to thank staff at the National Library of New Zealand (Wellington) for supplying archival material about Comet C/1910 A1.

Finally, this paper forms part of the first author's PhD project on historical aspects of New Zealand cometary astronomy, and he is grateful to the University of Southern Queensland for providing travel funding and other research support.

**John Drummond** became fixated with astronomy at the age of ten when his mother pointed out the Pot in Orion to him. From that moment on he was hooked on the Universe. Joining the Junior Section of the local Gisborne Astronomical Society not long after, John would regularly do group meteor watches, telescope viewing and listen to astronomy talks. He also developed an interest in photography, and it was not long before he combined these two interests and began astrophotography. John's photographs have been used in many overseas books and magazines—and were used on two New Zealand stamps. He was the Director of the Royal Astronomical Society of New Zealand's Astrophotography Section for thirteen years until 2018. He is currently the Director of the Society's Comet Section.

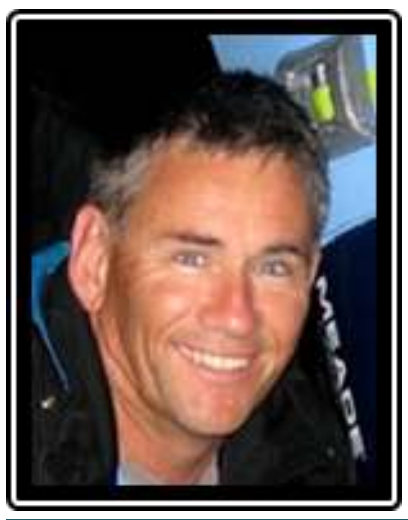

John lives about 10km west of Gisborne, on the east coast of the North Island of New Zealand, and has a range of telescopes up to 0.5 metres in aperture at his Possum Observatory. He regularly images with these telescopes and CCDs, carries out astrometry of comets, asteroids and NEOs, and sends his observations to the IAU Minor Planet Center. He has also confirmed several comets and co-discovered about 20 exoplanets in collaboration with the Ohio State University. He runs Gisborne Astro Tours from his observatories. John has authored or co-authored more than 60 research papers, many of the more recent ones being on the history of New Zealand astronomy.

John is a Past President and Past Secretary of the Royal Astronomical Society of New Zealand, and in 2019 he was made a Fellow of the Society. In 2016 he was awarded an MSc (Astronomy) by Swinburne University in Melbourne (Australia), and currently he is researching aspects of the history of cometary astronomy in New Zealand as a part-time off-campus internet-based PhD student affiliated with the Centre for Astrophysics at the University of Southern Queensland (Australia). His supervisors are the co-authors of this paper.

When he is not doing astronomy, John is a secondary school science teacher. He also enjoys surfing the great waves of Gisborne and pottering around on his small farm tending to his sheep.

**Professor Wayne Orchiston** was born in Auckland (New Zealand) in 1943, and has BA First Class Honours and PhD degrees from the University of Sydney. Currently, he is employed by the University of Science and Technology of China in Hefei as the Co-editor of the *Journal of Astronomical History and Heritage*. He is also an Adjunct Professor of Astronomy at the Centre for Astrophysics at the University of Southern Queensland (USQ) in Toowoomba, Australia. Formerly, Wayne worked at observatories, research institutes and universities in Australia, New Zealand and Thailand.

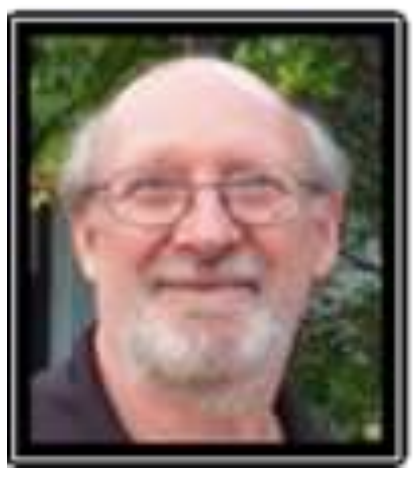

Over the past two decades Wayne has supervised more than 35 Master of Astronomy and PhD history of astronomy research projects through three different Australian universities.

Wayne has wide-ranging research interests and more than 500 publications, mainly about historic transits of Venus; historic solar eclipses; historic telescopes and observatories; the emergence of astrophysics in Asia and Oceania; the history of cometary and meteor astronomy; the astronomy of James Cook's three voyages to the Pacific; amateur astronomy and the amateur–professional interface; the history of meteoritics; Indian, Southeast Asian and Māori ethnoastronomy; and the history of radio astronomy in Australia, France, India, Japan, New Zealand and the USA.

Recent books by Wayne include *Exploring the History of New Zealand Astronomy …* (2016, Springer); *John Tebbutt: Rebuilding and Strengthening the Foundations of Australian Astronomy* (2017, Springer); *The Emergence of Astrophysics in Asia …* (2017, Springer, co-edited by Tsuko Nakamura); *Exploring the History of Southeast Asian Astronomy …* (2021, Springer, co-edited by Mayank Vahia) and *Golden Years of Australian Radio Astronomy: An Illustrated History* (2021, Springer, co-authored by Peter Robertson and Woody Sullivan). His latest book, *Histoire de la Radioastronomie Française*, co-authored by James Lequeux, will be published shortly. Wayne has also edited or co-edited a succession of conference proceedings.

Since 1985 Wayne has been a member of the IAU, and he is a former President of Commission C3 (History of Astronomy). In 2003 he founded the IAU's Historical Radio Astronomy Working Group, and is the current Radio Astronomy Subject Editor for the Third Edition of Springer's *Biographical Encyclopedia of Astronomers*. He also founded the IAU Working Group on Historic Transits of Venus, and is the founding Director of the large and dynamic Historical Section of the Royal Astronomical Society of New Zealand. Wayne is also an Editor of Springer's book series on Cultural and Historical Astronomy. In 2014 he founded the History & Heritage Working Group of the Southeast Asian Astronomy Network and ran this successfully until 2024. In the process he organised three different conferences on SE Asian Astronomical History; some of the presented papers ended up in the aforementioned Southeast Asian Springer book. Since 2004 Wayne has also served on the Executive Committee of the ICOA series of Asian–Oceanic conferences, and currently he is a Councillor of the Royal Astronomical Society of New Zealand

In 1998 Wayne Orchiston and John Perdrix co-founded the *Journal of Astronomical History and Heritage*. After John's death, Wayne was the Managing Editor until 31 July 2022 when he passed ownership of the journal to the University of Science and Technology of China. In 2013 the IAU named minor planet 48471 'Orchiston', and in 2019 former PhD student Stella Cottam, and Wayne, were awarded the Donald E.Osterbrock Prize by the American Astronomical Society for their 2015 Springer book, *Eclipses, Transits and Comets of the Nineteenth Century* … In 2023 he was elected an Honorary Member of the Royal Astronomical Society of New Zealand, and two of his former doctoral students edited the following Festschrift in his honour: Gullberg, S., and Robertson, P. (eds.), 2023. *Essays in Astronomical History and Heritage: A Tribute to Wayne Orchiston on His 80th Birthday* (Springer, 2023). In January 2024 the American Astronomical Society also awarded Wayne their LeRoy E. Doggett Prize for lifetime contributions to history of astronomy.

Wayne and Darunee Lingling Orchiston live in a quiet village near Chiang Mai in northern Thailand. When not involved in astronomy Wayne particularly enjoys following Australian and New Zealand athletics and Formula 1 and Indycar racing.

**Dr. Carolyn Brown** was born in Toowoomba, Queensland in Australia and has been interested in astronomy since she was four years old. After watching Stephen Hawking's "A Brief History of Time" in 1991, she proclaimed that she would become an astrophysicist. Which she did. Undertaking degrees in physics and astronomy at the University of Southern Queensland, Carolyn took on part-time work in the University's Physics and Astronomy Department while she completed her PhD. In 2009, Carolyn became a full-time academic in Physics and Engineering at the University of Southern Queensland while continuing to follow her passion in astronomy through research and outreach programs.

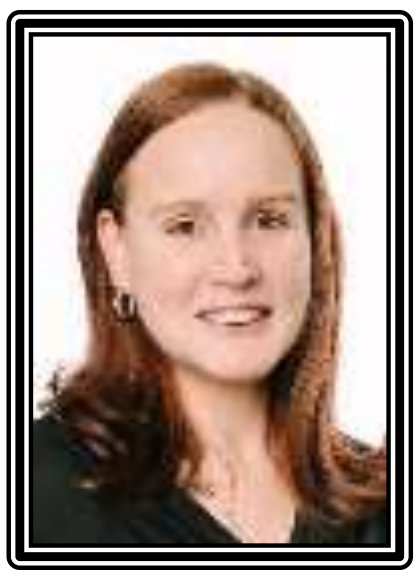

As Carolyn's career progressed, she found her love of teaching grew beyond her love of research and began to concentrate on educating the new generation in their passion for astronomy, like others had done for her. Taking on a primarily teaching role at the University, Carolyn continued to foster the passion for astronomy in her students, especially her postgraduate students, until one day, convinced by one of her Doctoral students, making the decision to move into the secondary school sector where she is hoping to make an impact on our youth to help drive them to follow their passions.

When not educating the next generation, Carolyn enjoys restoring classic Australian muscle cars where she can put her theoretical physics and engineering knowledge into a practical (and fast) application. Now all she needs is a bigger shed, with an observatory on the roof.

**Professor Jonathan (Jonti) Horner** was born in Wakefield, Yorkshire, in the United Kingdom, in 1978. At the age of five he saw part of an episode of "The Sky At Night" and became hooked on all things astronomical. He joined his local astronomical society—the West Yorkshire Astronomical Society (WYAS)—at the age of eight and began giving regular talks and writing for the society's journal, *Phobos*, by the age of ten. Thanks to the support and advice he received from both the members and the guest speakers at WYAS, Jonti moved to the University of Durham in 1996 to study for an undergraduate Masters' degree in Physics and Astronomy. After a summer project working at Armagh Observatory in 1999, with Professor Mark Bailey, Jonti moved to the University of Oxford in 2000, to begin work on his doctoral studies. His DPhil was conferred in 2004 for a thesis titled "The Behaviour of Small Bodies in the Outer Solar System." He spent time as a Postdoctoral Research Fellow at the University of Bern (Switzerland) and the Open University (UK), and a year as a teaching-only Fellow at Durham University (UK) before moving to Australia in 2010, to take up a Postdoctoral Fellowship at the University of New South Wales in Sydney. Finally, in 2014, Jonti moved to take up a position as Vice-Chancellor's Senior Research Fellow at the University of Southern Queensland in Too-woomba, where he remains to this day.

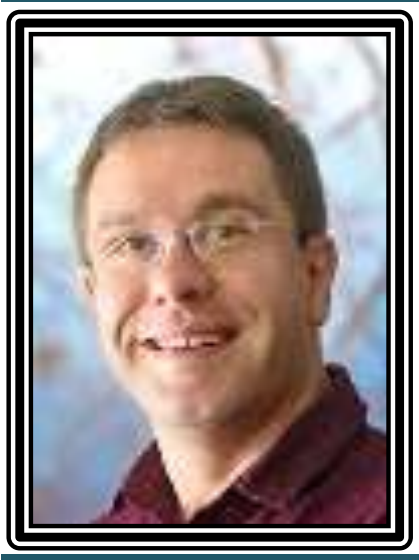

Jonti has a diverse range of research interests and has published more than 230 papers. His primary research focusses through his career have been the study of the Solar System's small bodies (particularly the Centaurs and planetary Trojans), the search for and characterisation of planets orbiting other stars (Exoplanets), and also the investigation of the various features that could render one planet more or less suitable as a target for the search for life beyond the Solar System. He is particularly proud of the work he did with his former mentor, Professor Barrie Jones, investigating the role that giant planets like Jupiter play in controlling the impact rate for terrestrial worlds—work that definitely shattered the long-held myth that Jupiter serves as Earth's celestial guardian and protector. Instead, Jupiter's role is more nuanced, with the giant planet actively throwing new objects to threaten the Earth with one hand whilst taking them away with the other. More recently, Jonti organised and led a lengthy review titled "Solar System Physics for Exoplanet Research", which has become a *de facto* textbook for undergraduate and Masters' courses around the world. His full publication list can be accessed through a library on the NASA ADS system, here: https://ui.adsabs.harvard.edu/public-libraries/YUefu-IISQSn4d6JVwKokQ

Jonti is a member of the Astronomical Society of Australia, and an ongoing Committee Member of the Astrobiology Society of the Britain. He is a past member of Australia's National Committee for Space and Radio Science, and serves as the Honorary President of the West Yorkshire Astronomical Society. He is an active and enthusiastic science communicator, regularly appearing in national and international media to discuss stories about planetary and exoplanetary science, and astrobiology. He has written more than one hundred articles for "The Conversation", and is currently serving as a guest presenter on the globally popular "SpaceNuts" podcast. Jonti's ORCID is 0000-0002-1160-7970.

When he is not working as a professional astronomer, Jonti is a keen photographer, both of wildlife and as an astrophotographer. He is also a member of the Toowoomba Philharmonic Society's chorus, and enjoys gaming with friends and family at his home in the Darling Downs.